\documentclass[fleqn,usenatbib]{mnras}

\usepackage{mathptmx}

\usepackage[T1]{fontenc}
\usepackage{ae,aecompl}

\usepackage{graphicx}	
\usepackage{amsmath}	
\usepackage{amssymb}	
\usepackage{wasysym}           
\usepackage{epstopdf}
\usepackage{mathrsfs}
\usepackage{natbib}
\usepackage{physymb}
\usepackage{color}

\title[Tidally-disrupted debris streams]{On the structure of tidally-disrupted stellar debris streams}

\author[Coughlin et al.]{
Eric R. Coughlin,$^{1,2}$\thanks{email: eric.coughlin@colorado.edu}
Chris Nixon,$^{3}$
Mitchell C. Begelman,$^{1,2}$
Philip J. Armitage,$^{1,2}$
\\
$^{1}$JILA, University of Colorado and National Institute of Standards and Technology, 440 UCB, Boulder, CO 80309-0440, USA \\
$^{2}$Department of Astrophysical and Planetary Sciences, University of Colorado, 391 UCB, Boulder, CO 80309-0391, USA\\
$^{3}$Department of Physics \& Astronomy, University of Leicester, Leicester LEI 7RH UK \\
}

\date{Accepted XXX. Received YYY; in original form ZZZ}

\pubyear{2016}

\begin{document}
\label{firstpage}
\pagerange{\pageref{firstpage}--\pageref{lastpage}}
\maketitle

\begin{abstract}
A tidal disruption event (TDE) -- when a star is destroyed by the immense gravitational field of a supermassive black hole -- transforms a star into a stream of tidally-shredded debris. The properties of this debris ultimately determine the observable signatures of TDEs. Here we derive a simple, self-similar solution for the velocity profile of the debris streams produced from TDEs, and show that this solution agrees extremely well with numerical results. Using this self-similar solution, we calculate an analytic, approximate expression for the radial density profile of the stream. We show that there is a critical adiabatic index that varies as a function of position along the stream above (below) which the stream is unstable (stable) to gravitational fragmentation. We also calculate the impact of heating and cooling on this stability criterion.
\end{abstract}




\section{Introduction}
When a star comes within a supermassive black hole's (SMBH) tidal radius $r_t = R_*(M_h/M_*)^{1/3}$, where $R_*$ is the stellar radius and $M_h$ and $M_*$ are the black hole and stellar masses, respectively, the tidal shear due to the hole overcomes the self-gravity of the star. The tidal force subsequently tears the star apart, with half of the torn stellar debris bound to the black hole, the other half unbound \citep{lac82, ree88}.

These tidal disruption events (TDEs) have been studied analytically, numerically, and observationally for nearly forty years. The earliest analytic studies showed that the rate of return of the bound material, $\dot{M}_{fb}$, should scale roughly as $\dot{M}_{fb} \propto t^{-5/3}$ at late times \citep{phi89}, and early numerical simulations supported this scaling \citep{eva89}. Dozens of putative TDEs have now been discovered (e.g., \citealt{kom99, gez08, bur11, cen12, bog14, mil15}; see \citealt{kom15} for a review of the observational status of TDEs), and the rates at which they have been discovered show tentative agreement with early estimates of the rate at which they should occur ($10^{-4}-10^{-5}$ per galaxy per year; \citealt{fra76, sto14}). 

\citet{cou15a} recently demonstrated numerically that the streams of tidally-stripped debris were gravitationally unstable when the adiabatic index of the gas was set to $\gamma = 5/3$, resulting in their fragmentation into bound clumps. These clumps, when accreted at discrete times, caused the late-time fallback rate to fluctuate about the $t^{-5/3}$ average (the minima being induced by the accretion of lower-density material in between clumps). \citet{cou16} then performed a more comprehensive numerical study of TDEs in which the pericenter distance was comparable to the tidal radius, varying the polytropic index of the gas. They found that, for stiffer equations of state ($\gamma \gtrsim 5/3$), collapse was induced sooner and was enhanced by a ``post-periapsis pancake'' that takes place soon after the disruption. 

Here we construct a model of the stream of debris produced from a TDE in an attempt to characterize its general properties. We first demonstrate the existence of a simple, self-similar solution for the radial velocity profile along the stream in Section 2, and we compare it to the results of past simulations and use it to determine the stream position as a function of radial distance and time. In Section 3 we show how the stream width varies with density and radial position. Section 4 provides approximate, analytic solutions for the density along the stream and we compare these expressions to numerically-computed values. We analyze the rough scaling of the density with time in Section 5, and we also demonstrate that there is a critical adiabatic index {}{that varies as a function of position along the stream and time} at which the stream fragments. {}{It is shown that, for an isentropic equation of state, the critical adiabatic index is equal to $\gamma_c = 5/3$ for the initial evolution of the stream, but at late times falls to as low as $\gamma_c = 4/3$ for the unbound portion of the stream}. Section 6 provides a discussion of our findings and considers the results of including other effects in our analytic treatment, and we summarize and conclude in Section 7.

\section{Velocity distribution}
\subsection{Self-similar velocity profile}
As was noted in \citet{cou16}, assuming that the tidal disruption takes place impulsively accurately approximates the radial positions of the gas parcels from a TDE when the impact parameter of the star is $\beta = r_t/r_p  =1$, where $r_p$ is the pericenter distance of the star. This impulse approximation states that the star is able to maintain hydrostatic equilibrium until it reaches the tidal radius, at which point it is ``destroyed," meaning that the  self-gravity and pressure of the resultant gas become negligible at this location and for any moment in time thereafter. Therefore, each gas parcel of the tidally-stripped debris moves solely under the influence of the gravitational field of the black hole.

When one applies the impulse approximation to the SMBHs we are considering here, the orbital eccentricities of the stream occupy a very narrow range of values centered around one. Therefore, the velocity along the stream is very nearly radial for times not long after disruption. Furthermore, as was also demonstrated in \citet{cou16}, the self-gravity of the stream keeps it narrowly confined in the $\phi$ and $\theta$ directions. As a first approximation we can thus let $\mathbf{v}(r,\theta,\phi,t) \simeq v_r(r,t)\hat{r}$, where $\mathbf{v}$ is the velocity vector of the material in the stream and $\hat{r}$ is the unit vector in the radial direction {}{pointing away from the SMBH}. If we further neglect the influence of pressure gradients and self-gravity in the $r$-momentum equation, both of which should be small in comparison to the gravitational field exerted by the hole, then the radial component of the momentum equation is

\begin{equation}
\frac{\partial{v_r}}{\partial{t}}+v_r\frac{\partial{v_r}}{\partial{r}} = -\frac{GM_h}{r^2} \label{rmom},
\end{equation}
where $M_h$ is the mass of the SMBH. 

As we noted above, the eccentricities of the orbits are very nearly one for the entire stream. Furthermore, the relevant timescale at any given distance $r$ from the hole is set by the local dynamical time, $\tau_d = r^{3/2}/\sqrt{2GM_h}$, so a reasonable assumption for the radial velocity is that it varies self-similarly according to

\begin{equation}
v_r = \sqrt{\frac{2GM_h}{r}}f(\xi) \label{vrsim},
\end{equation}
where

\begin{equation}
\xi = \frac{\sqrt{2GM_h}t}{r^{3/2}}
\end{equation}
is time normalized to the dynamical time at radius $r$. If we insert this ansatz into equation \eqref{rmom}, we find the following self-similar equation to be solved for $f$:

\begin{equation}
f' = \frac{f^2-1}{2-3\xi{f}} \label{fss},
\end{equation}
where a prime denotes differentiation with respect to $\xi$.

In addition to this equation, one must also impose a boundary condition on $f$ in order to obtain the full solution for the velocity profile. To do so, recall that the specific energies of the stream are narrowly distributed about zero and are

\begin{equation}
\epsilon = \frac{v_r^2}{2}-\frac{GM_h}{r} = \frac{GM_h}{r}\left(f^2-1\right), \label{energy}
\end{equation}
which shows that $f$ is narrowly confined to 1 initially. Investigating equation \eqref{fss}, the location at which $f = 1$ will correspond to $f' = 0$ if $\xi \neq 2/3$; however, if $f' = 0$, then it is easy to see by taking more derivatives of equation \eqref{fss} that every higher derivative of $f$ is also zero, i.e., $f(\xi) = 1$ is the solution if $f'=0$ when $f = 1$. Since we we require the energies of the gas parcels to be distributed about, but not exactly equal to, zero, we see that the point $f = 1$ must coincide with $\xi = 2/3$. Therefore, our boundary condition on $f$ is $f(2/3) = 1$, which we note corresponds to a critical point in equation \eqref{fss}\footnote{The derivative at this point for the non-trivial solution, it can be shown, is $f'(2/3) = -5/2$.}.

\begin{figure}
   \centering
   \includegraphics[width=3.4in]{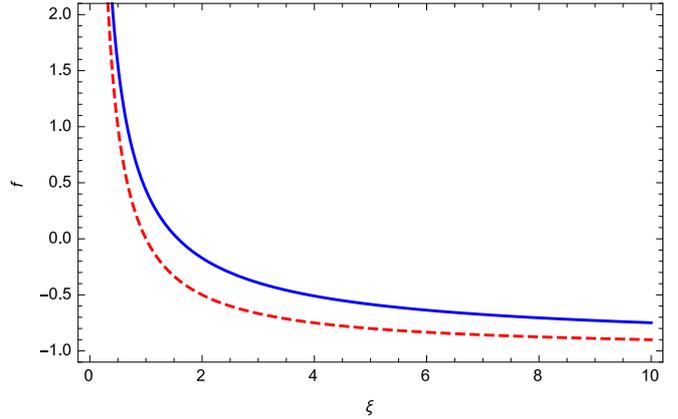} 
   \caption{The solution to equation \eqref{fss} passing through the critical point $f(2/3)=1$ (blue, solid curve). The approximate solution given by $f = -1+1/\xi$, which matches the asymptotic limits of the true solution, is shown by the red, dashed, curve. }
   \label{fig:fexapp}
\end{figure}

In further support of the fact that $f(2/3) = 1$ is the only boundary condition that can describe the debris stream, we note that equation \eqref{fss} can be integrated exactly to give

\begin{equation}
\frac{C+f\sqrt{f^2-1}-\ln\left(f+\sqrt{f^2-1}\right)}{\left(f^2-1\right)^{3/2}} = \xi,
\end{equation}
where $C$ is a constant of integration. This equation cannot be solved analytically to isolate $f(\xi)$; however, if we set $f = 1$, the left-hand side of this equation is singular \emph{unless} $C=0$. The only solution for $v_r$ that smoothly passes through marginally-bound portion of the stream must therefore have $C =0$. If we adopt this value of $C$ and use L'Hospital's rule to determine the limit of $f \rightarrow 1$, we find, as we expect, that $\xi \rightarrow 2/3$.

Figure \ref{fig:fexapp} shows this critical solution by the blue, solid curve. For contrast, the red, dashed curve shows the approximate solution $f \simeq -1+1/\xi$, which aids in seeing the asymptotic behavior of the true function. 

\begin{figure} 
   \centering
   \includegraphics[width=3.4in]{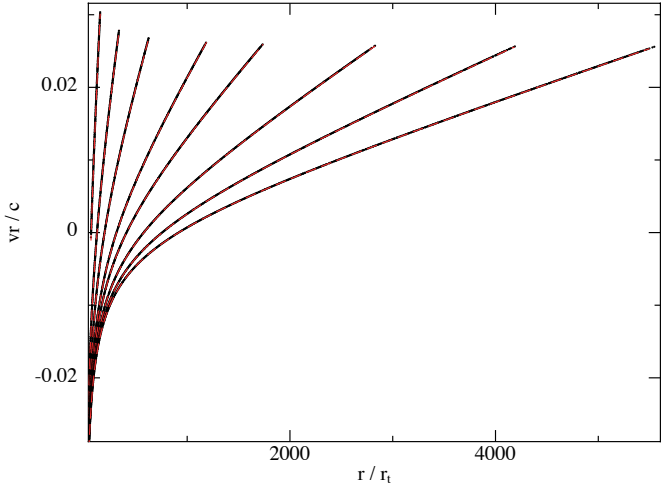} 
   \caption{A comparison between the self-similar (red, dashed curves) and numerical (black points) radial velocity profiles, both normalized to the speed of light, for the debris stream produced by the disruption of a solar-like star by a $10^6M_{\astrosun}$ black hole. The numerical solution was taken from \citet{cou15a}{}{, who used the smoothed-particle hydrodynamics (SPH) code {\sc{phantom}} \citep{pri10} to destroy a solar-like star by a $10^6M_{\odot}$ SMBH}. The different curves represent the stream at different times, with each curve corresponding to, from left to right, 11, 30, 60, 120, 180, 300, 451, and 601 days since disruption. }
   \label{fig:vcomps}
\end{figure}

In order to assess the validity of the self-similar solution in matching the radial velocity profile of tidally-disrupted debris streams, Figure \ref{fig:vcomps} compares the self-similar solution for $v_r$ (Equation \ref{vrsim}) to the numerical simulation of \citet{cou15a}. In this simulation a solar-type star (one with a solar mass and solar radius) was destroyed by a $10^6M_{\astrosun}$ SMBH {}{using the smoothed-particle hydrodynamics (SPH) code {\sc{phantom}} \citep{pri10} with $10^6$ particles}; the pericenter distance of the stellar progenitor was equal to the tidal radius, the gas followed an adiabatic, $\gamma = 5/3$ equation of state, and pressure and self-gravity were included at all stages of the simulation (we refer the interested reader to \citet{cou15a} for more details). The different curves in Figure \ref{fig:vcomps} correspond to the stream of debris at different times, with the earliest (farthest left on the plot) being roughly two weeks after disruption, the latest (occupying the largest radial extent) at almost two years (the Figure caption gives the precise times). This Figure shows that the analytic, self-similar solution agrees extraordinarily well with the numerical solution.

\begin{figure}
   \centering
   \includegraphics[width=3.4in, height=2.4 in]{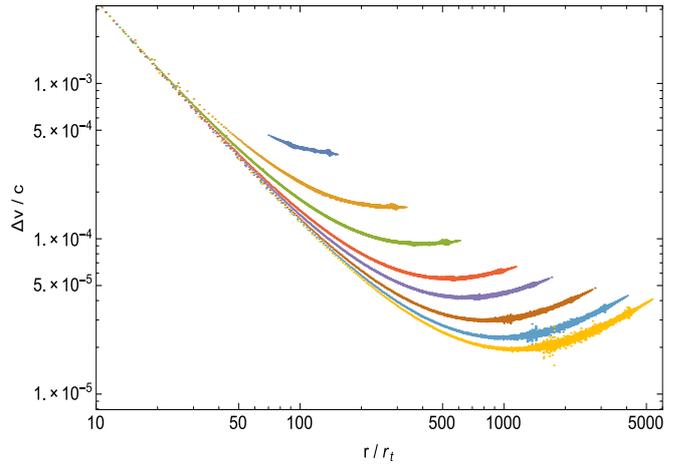} 
   \caption{A log-log plot of the difference in velocity normalized to the speed of light, i.e., $\Delta{v}/c \equiv (v_{n} - v_{ss})/c$, where $v_n$ is the numerically-obtained radial velocity and $v_{ss}$ is the analytic solution (equation \ref{vrsim}), for the same times chosen in Figure \ref{fig:vcomps} (i.e., the dark blue set of points is 11 days after disruption, the light yellow set of points is 601 days after disruption, and each set of points in between corresponds to the appropriate curve in Figure \ref{fig:vcomps}). As discussed in Section 6.1, the small discrepancies shown here can plausibly be attributed to the neglect of angular momentum in the self-similar solution.}
   \label{fig:deltavplots}
\end{figure}

Figure \ref{fig:deltavplots} shows the difference in velocity, $\Delta{v}$, between the numerically-obtained radial velocity profile and the analytic velocity profile, i.e., $\Delta{v} = v_n - v_{ss}$, $v_n$ being the numerical solution, $v_{ss}$ the analytic solution (equation \ref{vrsim}), normalized to the speed of light. Each set of points corresponds to the time chosen in Figure \ref{fig:vcomps}. The small differences illustrated by this plot may be due to the neglect of angular momentum in the self-similar solution (see Section 6.1).

\subsection{Radial positions}
With the function $f$ completely determined, the radial positions of the gas parcels comprising the stream can be found by solving the differential equation

\begin{equation}
\dot{r}_i = \sqrt{\frac{2GM_h}{r_i}}f(\xi_i) \label{rdot},
\end{equation}
where $r_i(t)$ is the radial position of gas parcel $i$, $\xi_i = \sqrt{2GM_h}t/r_i^{3/2}$ is the self-similar variable of gas parcel $i$, and dots denote differentiation with respect to time. 

Because $f$ is not analytic, this equation cannot be integrated exactly to give $r_i(t)$. However, we note that there is an exact solution for the marginally-bound material: 

\begin{equation}
r_m = \left(\frac{3}{2}\sqrt{2GM_h}t\right)^{2/3} \label{rm}.
\end{equation}
As a check, differentiating this equation with respect to time gives $\dot{r}_i = \sqrt{2GM_h/r}$, which, from equation \eqref{rdot}, demands $f(\xi_m) = 1$. On the other hand, $\xi_m = \sqrt{2GM_h}t/r_m^{3/2} = 2/3$, and since our boundary condition is $f(2/3) = 1$, we see that equation \eqref{rm} does indeed solve equation \eqref{rdot}. 

To determine the remaining positions of the gas parcels, we will apply an argument similar to the impulse approximation: at some time $t_m$, the stream is closely confined to the position of the marginally-bound segment of the stream, with its radial extent equal to the diameter of the disrupted star. We will thus let

\begin{equation}
r_i(t_m) = r_m(t_m)+\delta{r}_i \label{ridelta},
\end{equation}
where $\delta{r}_i =  \mu_i{R_*} \ll r_m(t_m)$ and $|\mu_i| < 1$. Furthermore, we know that the tidal force of the black hole imparts a spread of energies to the debris of \citep{lac82, ree88, cou16}

\begin{equation}
{\epsilon}_i = \frac{GM_hR_*}{r_t^2}\mu_i \label{eps},
\end{equation}
where $\mu_i$ is the dimensionless distance of gas parcel $i$ from the center of mass of the star at the time of disruption. However, from equation \eqref{energy} we have that the energies of the gas parcels at time $t_m$ are

\begin{equation}
\epsilon_i = \frac{GM_h}{r_i}\left(f^2-1\right),
\end{equation}
where the entire right-hand side is evaluated at $t_m$, and since the stream is narrowly confined to the position of the marginally-bound parcel, we have

\begin{equation}
f \simeq 1-\frac{5}{2}\left(\xi-\frac{2}{3}\right) \label{fapp}.
\end{equation}
This equation is just a Taylor expansion of $f$ about the point $\xi = 2/3$, and we used the fact that $f'(2/3) = -5/2$ (see footnote above). We therefore have that

\begin{equation}
\epsilon_i = -\frac{5GM_h}{r_i}\left(\xi-\frac{2}{3}\right),
\end{equation}
and using equations \eqref{ridelta} and \eqref{eps} and keeping only first order terms in $\delta{r}_i$, we find that the positions of the gas parcels satisfy

\begin{equation}
\delta{r}_i = \frac{1}{5}\frac{r_m^2}{r_t^2}R_*\mu_i.
\end{equation}
Setting $\delta{r}_i = \mu_iR_*$, we see that the time at which the energies are ``frozen in" corresponds to $r_m = \sqrt{5}r_t$ in the self-similar approximation. This shows that, if we adopt our self-similar function for the velocities, we cannot impose the impulse approximation exactly, which would amount to setting the center of the star at $r_t$ at $t=0$ with the radial extent of the stream equal to $2R_*$. This occurs because, by specifying the positions of the gas parcels at some time $t_0$, we automatically know their velocities via equation \eqref{rdot} and, consequently, their energies. Therefore, we do not expect that the energies will automatically correspond to the correct values (equation \ref{eps}) by enforcing $r_i(t_0) = r_t+\mu_iR_*$, and indeed we see that we must have $r_i(t_0) = \sqrt{5}r_t+\mu_iR_*$ to correctly match the energies.  

With the initial conditions $r_i = r_m(t_m) + \mu_iR_*$, $r_m(t_m) = \sqrt{5}r_t$, and $t_m$ given by equation \eqref{rm} with $r_m = \sqrt{5}r_t$, we can numerically integrate equation \eqref{rdot} to solve for the positions of the gas parcels as a function of time. However, we can obtain an approximate expression for the initial evolution of the debris stream by letting 

\begin{equation}
r(\mu, t) \simeq r_m(t)+\mu{r_1(t)} +\mathcal{O}(\mu^2) \label{rmuapp},
\end{equation}
where we have let $\mu_i \rightarrow \mu$ become a continuous variable and $r_1$ is a small correction (on the order of $R_*/r_t$) to the marginally-bound orbit $r_m(t)$ (note that this is just a Taylor expansion of $r(\mu,t)$ about $\mu$, truncated at first order). If we now insert equation \eqref{rmuapp} into equation \eqref{rdot}, use equation \eqref{fapp} for the function $f$, and keep only terms to first order in $\mu{r_1}$, performing a bit of algebra reveals that

\begin{equation}
\frac{\dot{r}_1}{r_1} = \frac{4}{3\,t} \label{r1app}.
\end{equation}
This shows that the correction to the positions of the material from the marginally-bound orbit, during the initial evolution of the debris stream, scales as $r_1 \propto t^{4/3}$. Using this expression for $r_1$ in equation \eqref{rmuapp}, this also shows that

\begin{equation}
\frac{\partial{r}}{\partial{\mu}}\bigg{|}_{\mu = 0} = r_1(t) \propto t^{4/3}.
\end{equation}
Note that this result is exact, as all of the higher-order $\mu$ corrections are zero at $\mu = 0$. We will use this result in Section 3.2. 

Finally, while equation \eqref{rmuapp} is useful for the early evolution of the debris, it breaks down once the orbits of the material start to deviate from $r_i\propto t^{2/3}$. We can obtain an approximate expression for the positions of the gas parcels, one that is roughly valid over the whole stream, by recalling that the function $f$ is reasonably well-approximated by the function

\begin{equation}
f \simeq -1+\frac{1}{\xi} \label{fapp1},
\end{equation}
as depicted in Figure \ref{fig:fexapp}. If we use this approximate form for $f$, then we can show that equation \eqref{rdot} can be solved for $r_i(t)$ to give

\begin{equation}
r_i(t) = \left(3\sqrt{2GM_h}t+A_it^{3/2}\right)^{2/3}, \label{rapp}
\end{equation}
where $A_i$ is a constant of integration particular to gas parcel $i$. We can determine these constants by, as above, requiring that the length of the stream is narrowly confined, initially, about the marginally-bound position, i.e., $r_i(t_m) = r_m(t_m)+\delta{r}_i$, with $\delta{r}_i = \mu_iR_*$. Doing so gives

\begin{equation}
A_i = \frac{9\sqrt{3}}{2}\frac{\left(2GM_h\right)^{3/4}R_*}{r_m^{7/4}}\mu_i \label{Aiofmu}.
\end{equation}
Unfortunately, the value of $r_m$ cannot be determined by requiring that the energies match those from the impulse approximation; this is because the marginally-bound orbit from equation \eqref{rapp} -- where $r \propto t^{2/3}$ -- occurs at $\xi = 1/3$, while from equations \eqref{energy} and \eqref{fapp} it occurs where $\xi = 1/2$. This discrepancy means that the energies do not correlate with the properties of the orbits, and arises ultimately from the fact that the boundary condition $f(2/3) = 1$ is not satisfied by equation \eqref{fapp1}. Therefore, in this case we will simply let $r_m = \sqrt{5}r_t$ -- the same value found above when the correct form of $f$ is used -- and note that the positions of the gas parcels from equation \eqref{rapp} will differ from the true values because of the erroneous energies. We thus find that the approximate positions of the gas parcels are

\begin{equation}
r_i(t) = \left(3\sqrt{2GM_h}t+\chi\mu_i{t}^{3/2}\right)^{2/3}, \label{riapp}
\end{equation}
where we defined

\begin{equation}
\chi \equiv \frac{9\sqrt{3}}{2}\frac{(2GM_h)^{3/4}R_*}{r_m^{7/4}}.
\end{equation}

\section{Stream width}
The previous section exploited the fact that, to a very good approximation, the radial positions of the gas parcels can be found by considering them as non-interacting test particles in the gravitational field of the hole; this notion is substantiated by Figure \ref{fig:vcomps}. However, this will not necessarily be the case for the transverse structure of the stream, as self-gravity and pressure can be important at both early and late times \citep{koc94, cou15a, cou16}. If we account for self-gravity and pressure, the transverse momentum equation becomes

\begin{equation}
\frac{\partial{v_s}}{\partial{t}}+v_r\frac{\partial{v_s}}{\partial{r}}+v_s\frac{\partial{v_s}}{\partial{s}}+\frac{1}{\rho}\frac{\partial{p}}{\partial{s}}= g_{sg}+g_{M,\perp} \label{smom},
\end{equation}
where $s$ is the cylindrical distance from the stream center, $v_s$ is the velocity in the $s$-direction, $g_{sg}$ is the force due to self-gravity in the $s$ direction, and $g_{M,\perp}$ is the gravitational force arising from the black hole in the $s$-direction. Poisson's equation can be written

\begin{equation}
\nabla\cdot{\mathbf{g_{sg}}} = -4\pi{G}\rho.
\end{equation}
If we assume that the radial dependence of $g_{sg}$ is small in comparison with its $s$-dependence, which should be a good approximation owing to the degree of symmetry of the stream about its center of mass and breaks down only near its radial extremities, then this equation can be integrated using Gauss' law to give

\begin{equation}
g_{sg} \simeq -2\pi{G}\rho{s}.
\end{equation}
The perpendicular component of the gravitational field of the hole can also be written down as

\begin{equation}
g_{M,\perp} \simeq -\frac{GM_h{s}}{r^3}. \label{bhshear}
\end{equation}
Substituting these equations into equation \eqref{smom} then gives

\begin{equation}
\frac{\partial{v_s}}{\partial{t}}+v_r\frac{\partial{v_s}}{\partial{r}}+v_s\frac{\partial{v_s}}{\partial{s}}+\frac{1}{\rho}\frac{\partial{p}}{\partial{s}} = -2\pi{G}\rho{s}-\frac{GM_hs}{r^3}. \label{smom2}
\end{equation}
\subsection{Quasi-hydrostatic width}
During the early evolution of the debris, the pressure and self-gravity of the material are high enough that the stream remains approximately in gravitational equilibrium in the transverse direction (approximately because the density is evolving with time and the planar motions can cause the stream to be slightly over-pressured, resulting in oscillations; \citealt{cou16}). In this limit we can balance the pressure and self-gravity terms in equation \eqref{smom2}, which gives

\begin{equation}
H_{eq}^2 \simeq \frac{p}{2\pi\rho^2{G}},
\end{equation}
Here we will assume that the pressure varies as

\begin{equation}
p = S(r,t)\rho^{\gamma} \label{pofrho},
\end{equation}
where $S(r,t)$ is related to the entropy of the gas. A perfect polytrope is obtained by setting $S(r,t)$ to a constant; however, the stream of debris will cool radiatively as it evolves, and will do so increasingly as the material becomes optically thin to the radiation released during recombinations. Also, recombinations can heat the gas when it is still optically thick \citep{kas10}, and shocks can likewise serve to heat the material if the star is on a deeply-plunging orbit. To model these thermodynamic aspects of the problem, we will therefore permit $S(r,t)$ to be a function of $r$ and $t$ that could, in principle, be determined from the gas energy equation if all of the effects controlling the internal energy of the gas were known. The equilibrium width is thus

\begin{equation}
H_{eq}^2 = \frac{S\rho^{\gamma-2}}{2\pi{}G}, \label{Hsg}
\end{equation}
which is the same scaling found by \citet{cou16}.

\subsection{Shear dominated}
Depending on the adiabatic index of the gas or the amount of heating and cooling, the density will fall off at such a rate that the tidal shear term on the right-hand side of equation \eqref{smom2} will overcome the self-gravity \citep{cou16}. The pressure gradient will also be insufficient to balance the tidal compression when this happens, meaning that the advective terms on the left-hand side of equation \eqref{smom2} must become non-negligible. In this limit equation \eqref{smom2} becomes

\begin{equation}
\frac{\partial{v_s}}{\partial{t}}+v_r\frac{\partial{v_s}}{\partial{r}}+v_s\frac{\partial{v_s}}{\partial{s}} = -\frac{GM_hs}{r^3}.
\end{equation}
We expect that the solution for $v_s$ will again scale self-similarly, as the dynamical time is the only relevant timescale and only the gravitational field of the hole serves to alter the velocity $v_s$. We find that indeed there is a self-similar solution, and it is given by

\begin{equation}
v_s = \frac{s}{r}\sqrt{\frac{2GM_h}{r}}f(\xi),
\end{equation}
where $f(\xi)$ is the {same function} that appears in the distribution of the radial velocity. Since $v_s = d{s_i}/dt$, where $s_i$ is the transverse position of gas parcel $i$, and $\sqrt{2GM_h/r}f(\xi) = dr_i/dt$, we find that the transverse extent of the stream evolves as

\begin{equation}
H_i(t) = \frac{H_{i,0}}{r_{i,0}}r_i(t), \label{Hex}
\end{equation}
where $H_{i,0}$ is the width of the stream at time $t_0$. When the stream is dominated by the tidal shear of the hole, this equation shows that the width of the stream simply scales in proportion to its radial position and the entire evolution proceeds self-similarly.

\subsection{Approximate, full solution}
The above two restricted solutions demonstrate that, when the self-gravity of the stream dominates over the tidal shear of the black hole, the width of the stream is set by hydrostatic balance. On the other hand, when the tidal force of the black hole overwhelms self-gravity, $H \propto r$ and the entire stream evolves self-similarly. 

Along the full radial extent of the debris, the extremities will typically be shear-dominated, while the density in the central region is high enough to dominate over the shear of the hole \citep{koc94, cou15a}. We therefore expect the scaling of $H$ to vary between the self-gravity-dominated limit and the shear-dominated limit as we traverse the extent of the stream. In between those two limits, both the shear terms and the self-gravity terms are important in equation \eqref{smom2}, and as a result we do not expect a simple solution for the velocity $v_s$.  

We can, however, obtain an approximate solution that interpolates between the self-gravity and shear-dominated extremes by simply letting $H$ be piecewise defined about the point where $2\pi\rho = M_h/r^3$. We thus have

\begin{equation}
H^2 = \begin{cases}
\frac{S\rho^{\gamma-2}}{2\pi{G}} \quad \text{ for } 2\pi\rho \ge \frac{M_h}{r^3} \\
\\
\frac{H_0^2}{r_0^2}r^2 \quad \text{ for } 2\pi\rho \le \frac{M_h}{r^3}
\end{cases}. \label{Hfull}
\end{equation}
To ensure the continuity of the solution, $H_{0}$ and $r_{0}$ are given by the self-gravitating solution where $2\pi\rho = M_h/r^3$. In the next section we will see how $\rho$ is related to $H$.

\section{Density}
In addition to knowing the geometry of the stream, we would also like to be able to infer its density. To do so, we note that the mass contained in some segment of the stream is

\begin{equation}
M = \int_{V(t)}\rho(V,t)dV,
\end{equation}
where $V(t)$ is the volume of the stream segment that is, in general, time-dependent. The total mass contained in the segment is, however, time-{independent}, so the differential equation to be solved for $\rho$ is

\begin{equation}
\frac{d}{d{t}}\bigg{(}\int_{V(t)}\rho(V,t)\,dV\bigg{)} = 0 \label{rhoeq1}.
\end{equation}

When the limits of integration in this equation are functions of time, using the fundamental theorem of calculus results in the continuity equation. However, if we can transform to a specific set of coordinates in which the volume element is time-\emph{independent}, then the resulting equation can be immediately inverted to solve for $\rho$. In particular, if we write the new set of coordinates collectively as $T$, then the equation for $\rho$ is simply

\begin{equation}
\rho\frac{\partial{V}}{\partial{T}} = \frac{\partial{M}}{\partial{T}},
\end{equation}
where $\partial{V}/\partial{T}$ is the Jacobian of the transformation between the physical ($V$) and time-independent ($T$) coordinates. Here $\partial{M}/\partial{T}$ is the differential amount of mass contained in the volume element $dT$. 

We can achieve the transformation $V \rightarrow T$ by choosing our physical coordinates as $(r,\,s,\,\varphi)$, where $r$ is the radial position of the center of the stream, $s$ is cylindrical distance measured perpendicular from $r$, and $\varphi$ is the angle swept out by $s$ as it revolves around $r$. In terms of these coordinates, the mass contained in any segment of the stream is

\begin{equation}
M = \int{\rho}\,s\,ds\,d\phi\,dr.
\end{equation}
If we further define the $H$-averaged density as

\begin{equation}
\pi{H^2}\bar\rho \equiv \int{\rho\,s\,ds\,d\phi},
\end{equation}
then the mass is given by

\begin{equation}
M = \pi\int_{r_0(t)}^{r_1(t)}H^2\rho\,dr,
\end{equation}
where we dropped the bar for ease of notation. By now transforming from $r \rightarrow \mu$, where $\mu$ is the dimensionless original position of the stream, we find (see also equation 2 of \citealt{cou15a})

\begin{equation}
\pi{}H^2\rho\frac{\partial{r}}{\partial{\mu}} = \frac{\partial{M}}{\partial{\mu}} \label{rho00}.
\end{equation}

In order to make more progress with this equation, we must determine the quantities $H(\rho,r)$, $\partial{r}/\partial\mu$ and $\partial{M}/\partial\mu$. As we saw in the previous section, $H$ depends on $\rho$ and $r$ in a non-trivial way, and will behave differently depending on whether the stream is shear dominated or self-gravity dominated. However, we noted that a full, approximate solution can be obtained by assuming that $H$ switches between its hydrostatic equilibrium value and self-similar expansion at the point where $2\pi\rho = M_h/r^3$, which gives equation \eqref{Hfull} for $H$. 

The most rigorous way of calculating $\partial{r}/\partial\mu$ is to do so numerically, i.e., calculate the positions of the gas parcels comprising the stream, $r(\mu,t)$, via equation \eqref{rdot} for a large number of $\mu$ and $t$, and at a specific time $t$ interpolate over all of the $\mu$ to determine $\partial{r}/\partial{\mu}$. However, we can also use equation \eqref{riapp}, which gives an approximate solution for the positions of the gas parcels as functions of $\mu$ and $t$, to obtain an analytic expression. Doing so gives

\begin{equation}
\frac{\partial{r}}{\partial{\mu}} = \frac{2}{3}\chi{t^{3/2}}r^{-1/2} \label{drdmu},
\end{equation}
which we note is approximately valid over the entire extent of the stream (this expression, however, gives $\partial{r}/\partial\mu|_{\mu = 0} \propto t^{7/6}$, in contrast to the exact value of $\partial{r}/\partial\mu|_{\mu = 0} \propto t^{4/3}$). 

\begin{figure*}
   \centering
   \includegraphics[width=0.49\textwidth]{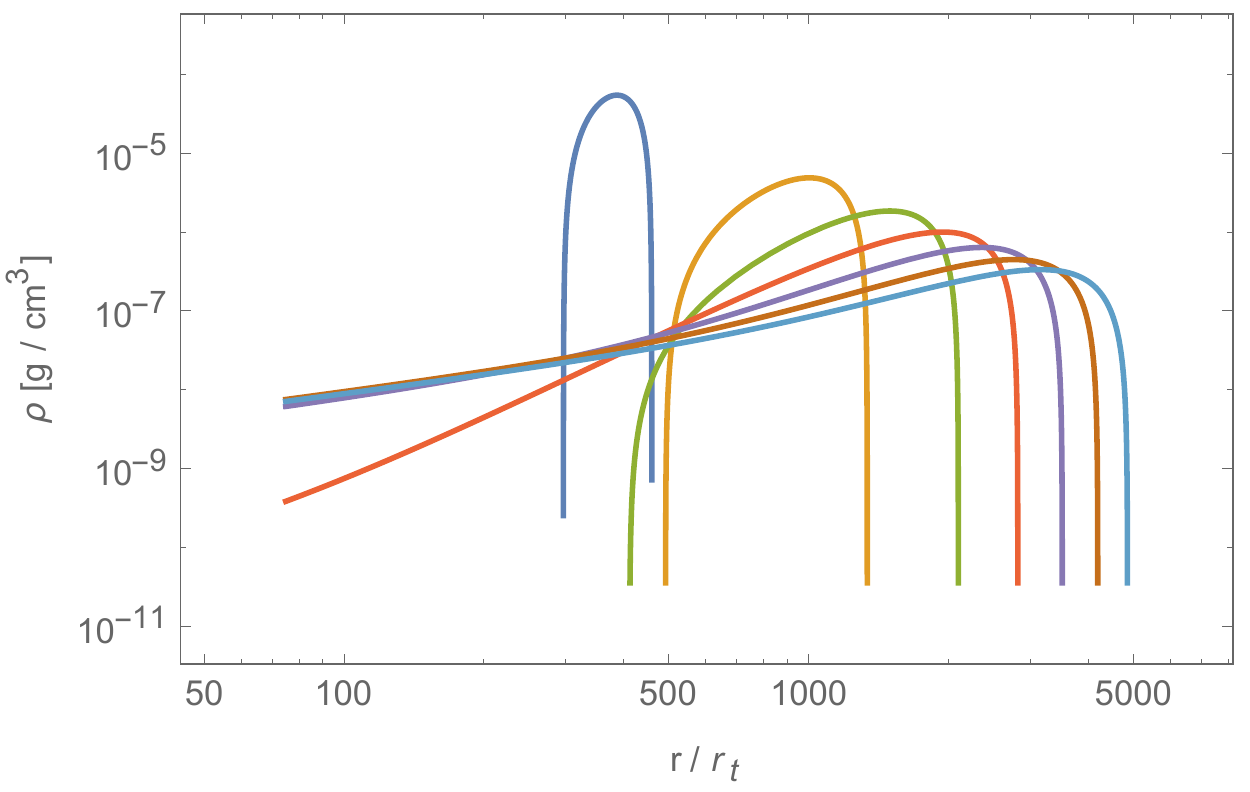} 
   \includegraphics[width=0.49\textwidth]{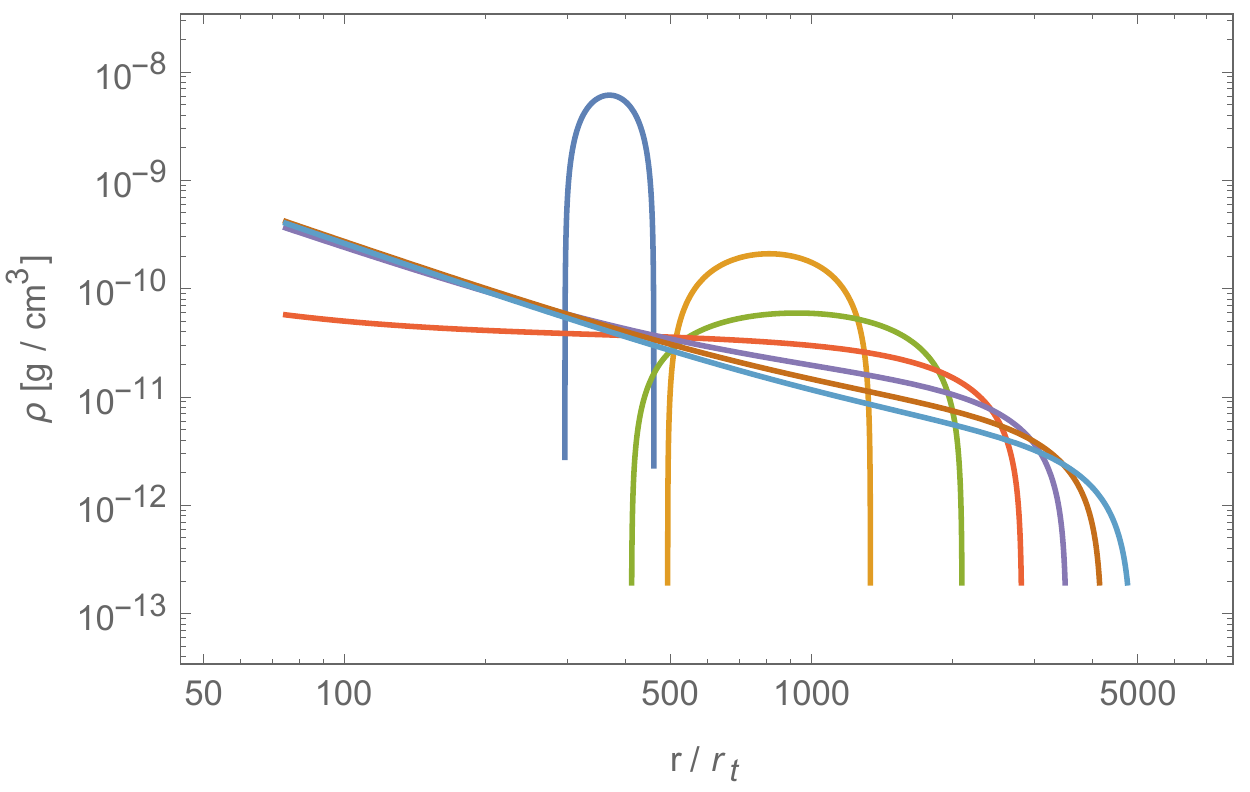} 
   \caption{The analytic solution for the density when self-gravity dominates the shear of the black hole (equation \ref{rhosg}), shown in the left panel, and when the shear dominates the self-gravity of the stream (equation \ref{rhosh}), shown in the right panel, as functions of $r$ for the disruption of a solar-like star by a $10^6M_{\odot}$ SMBH. We assumed isentropic gas with $\gamma = 5/3$ and entropy calculated for solar-like parameters ($S = 2.48\times10^{14}$ [cgs]; \citealt{han04}). The different colors correspond to different times, with the earliest time being $2500\times{}r_{t}^{3/2}/\sqrt{2GM_h} \simeq 32$ days after disruption (left-most, dark blue curve), the latest being $47500\times{}r_t^{3/2}/\sqrt{2GM_h} \simeq 620$ days after disruption (right-most, light blue curve), and each curve differing from the previous one by $7500\times{}r_t^{3/2}/\sqrt{2GM_h} \simeq 98$ days.}
   \label{fig:rhoplots}
\end{figure*}

\begin{figure*}
   \centering
   \includegraphics[width=0.495\textwidth]{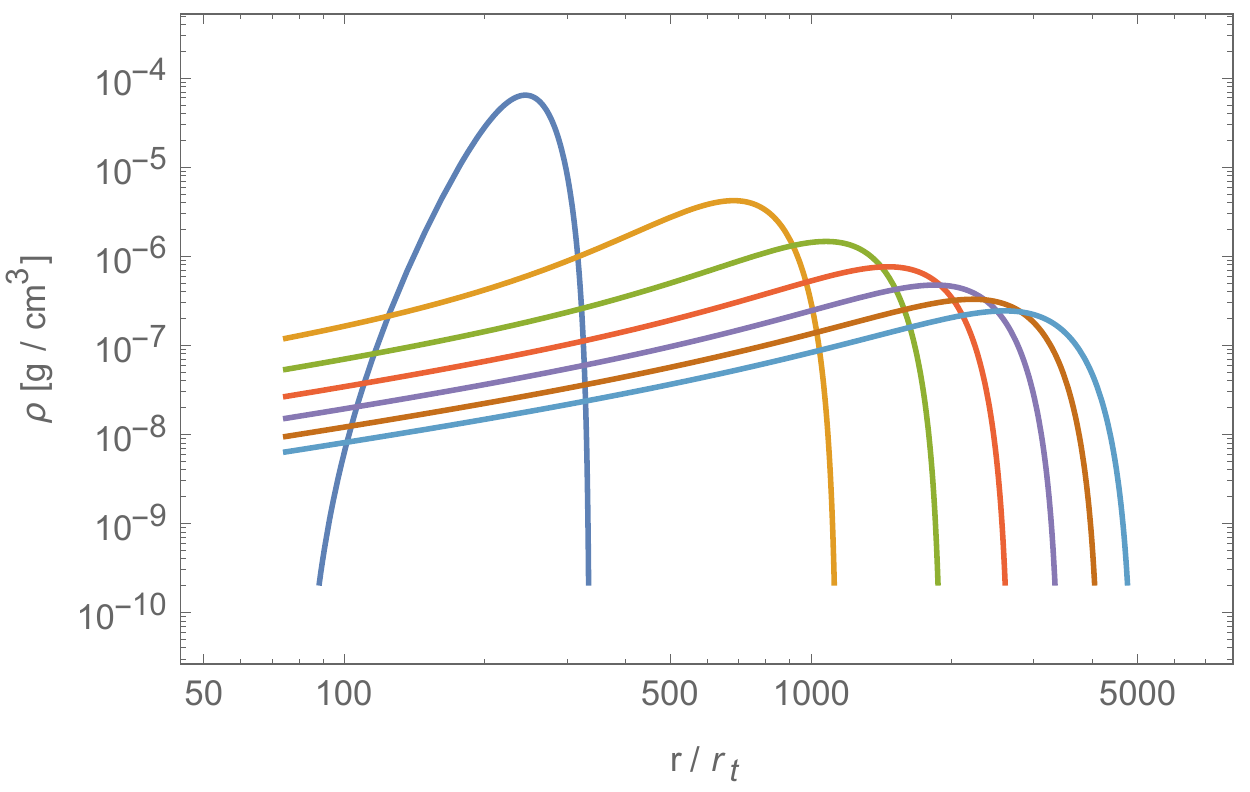} 
   \includegraphics[width=0.495\textwidth]{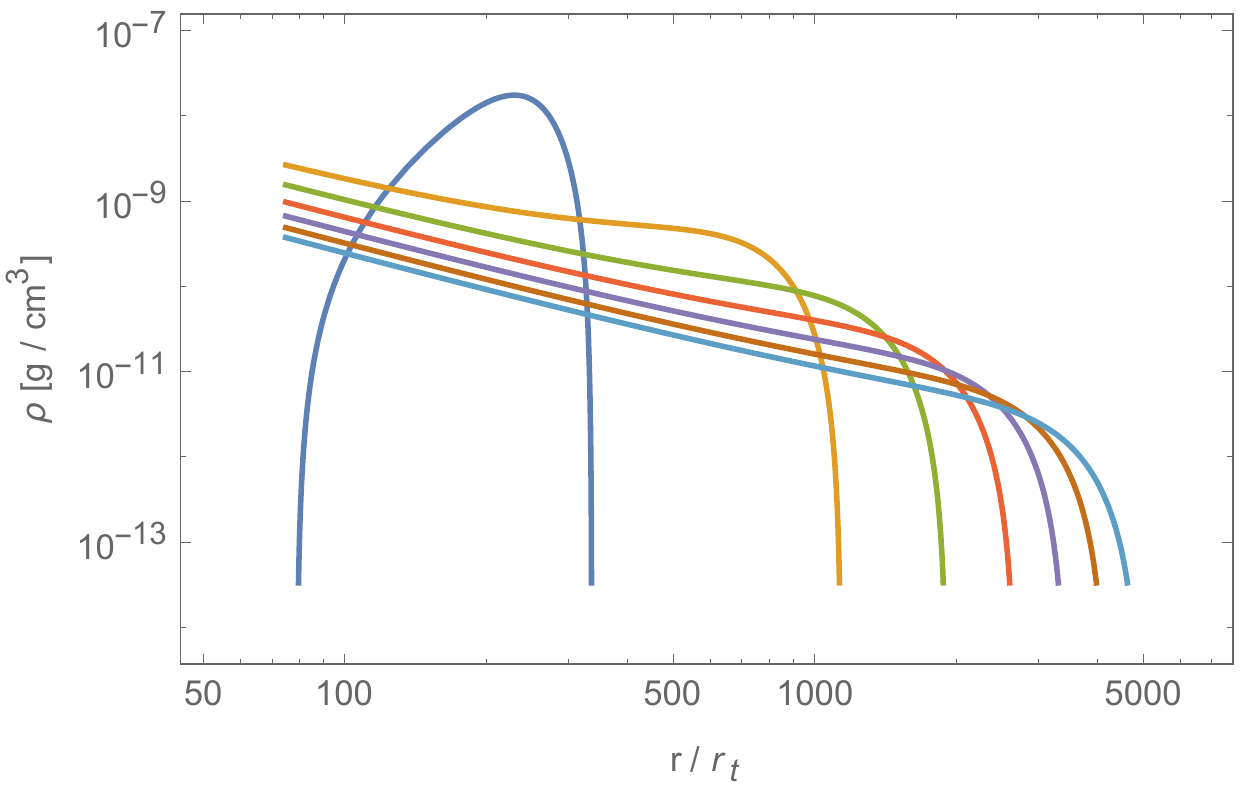} 
   \caption{The numerical solution for the density when self-gravity dominates the shear of the black hole (the solution to equation \ref{rho00} with $H$ given by equation \eqref{Hsg}, $\partial{r}/\partial\mu$ calculated numerically, and $\partial{M}/\partial\mu$ calculated numerically from equation \eqref{dmdmuex} with $n = 1.5$), shown in the left panel, and when the shear dominates the self-gravity of the stream (the solution to equation \ref{rho00} with $H$ given by equation \eqref{Hex}, $\partial{r}/\partial\mu$ calculated numerically, and $\partial{M}/\partial\mu$ calculated numerically from equation \eqref{dmdmuex} with $n = 1.5$), for the same parameters and times chosen in Figure \ref{fig:rhoplots}. }
   \label{fig:rhoplotsex}
\end{figure*}

Finally, the function $\partial{M}/\partial{\mu}$ can be determined by using the impulse approximation and considering the star at the time of disruption. This assumption then gives (see \citet{lod09}, \citet{cou14a} and \citet{cou16} for details)

\begin{equation}
\frac{\partial{M}}{\partial{\mu}} = \frac{1}{2}M_*\xi_1\frac{\int_{|\mu|\xi_1}^{\xi_1}\Theta(\xi)^n\xi{}d\xi}{\int_0^{\xi_1}\Theta(\xi)^n\xi^2d\xi}, \label{dmdmuex}
\end{equation}
where $\Theta(\xi)$ is the solution to the Lane-Emden equation and $\xi$ is the dimensionless radius of the polytrope \citep{han04}. A slight difficulty arises from the fact that, if the energies of the gas parcels using our self-similar prescription are to match those from the true impulse approximation (the one that accounts for the finite angular momentum of the material), then the ``time of disruption'' -- the point at which the energies are frozen in -- corresponds to when the marginally-bound segment of the stream is at a distance of $r_m = \sqrt{5}r_t$, not at $r_t$ (see Section 2.2). This means that using equation \eqref{dmdmuex} with the self-similar solution will overestimate the density, as we expect that the stream will have stretched by some factor by the time it reaches the distance $\sqrt{5}r_t$ and correspondingly decreased the density.

Since the functions $\Theta(\xi)$ are only analytic for a select few values of $n$, equation \eqref{dmdmuex} cannot, in general, be simplified further. However, we can obtain an approximate solution by letting $\Theta(\xi) \simeq 1$, which is valid in the inner regions of all polytropes. We can therefore write

\begin{equation}
\frac{\partial{M}}{\partial{\mu}} \simeq \frac{3}{4}M_*\left(1-\mu^2\right). \label{dmdmu}
\end{equation}

Because $H$ switches between its equilibrium value and self-similar expansion along the length of the stream, equation \eqref{rho00}, even with the approximate, analytic expressions for $\partial{r}/\partial\mu$ and $\partial{M}/\partial{\mu}$, cannot be rearranged to write $\rho$ in closed-form over the entire extent of the stream. However, in the limit that self-gravity or the tidal shear dominates, we can solve the equation exactly. Inserting equations \eqref{drdmu} and \eqref{dmdmu} into equation \eqref{rho00} and inverting the function \eqref{rapp} to solve for $\mu(r,t)$, we find that these limits correspond to

\begin{equation}
\rho_{sg}^{\gamma-1} = B_{sg}\frac{r^{1/2}}{t^{3/2}} \left(1-\frac{r^3}{\chi^2t^3}\left(1-\frac{3\sqrt{2GM_h}t}{r^{3/2}}\right)^2\right) \label{rhosg}
\end{equation}

\begin{equation}
\rho_{sh} = B_{sh}\frac{r^{-3/2}}{t^{3/2}}\left(1-\frac{r^3}{\chi^2t^{3}}\left(1-\frac{3\sqrt{2GM_h}t}{r^{3/2}}\right)^2\right) \label{rhosh}
\end{equation}
where $\rho_{sg}$ and $\rho_{sh}$ refer to the density in the self-gravity and shear-dominated cases, respectively, and for compactness we defined

\begin{equation}
B_{sg} \equiv \frac{9GM_*}{4S\chi} \label{rho0sg}
\end{equation}
and

\begin{equation}
B_{sh} = \frac{45M_*}{8\pi\chi}\left(\frac{M_h}{M_*}\right)^{2/3}. \label{rho0sh}
\end{equation}
For the shear-dominated solution, we let $H_0 = R_*$ when $r = \sqrt{5}r_t$, which is consistent with the impulse approximation under our self-similar prescription as we saw in Section 2.2.

Figure \ref{fig:rhoplots} shows these two solutions, the left panel the self-gravity dominated case, the right panel being shear dominated, when $M_h = 10^6M_{\odot}$, $R_* = 1R_{\odot}$, $M_* = 1M_{\odot}$, $\gamma = 5/3$, and the gas is isentropic with $S = 2.48\times10^{14}$ [cgs], which is the value appropriate for a Sun-like star \citep{han04}. The different colors correspond to different times since disruption, with the earliest time being $2500\times{}r_{t}^{3/2}/\sqrt{2GM_h} \simeq 32$ days after disruption (left-most, dark blue curve), the latest being $47500\times{}r_t^{3/2}/\sqrt{2GM_h} \simeq 620$ days after disruption (right-most, light blue curve), and each curve differing from the previous one by $7500\times{}r_t^{3/2}/\sqrt{2GM_h} \simeq 98$ days.We see that the density structure differs substantially between the two analytic cases, retaining an ``inverted'' profile when the stream is self-gravitating (the density increases as a function of $r$), but quickly assuming a density profile that decreases monotonically for the shear-dominated case. 

Figure \ref{fig:rhoplotsex} shows the numerical solution to equation \eqref{rho00} when self-gravity dominates (left-hand panel) and when shear dominates (right-hand panel). For these solutions, we calculated $\partial{r}/\partial\mu$ directly by numerically integrating equation \eqref{rdot} for a large number of $\mu$ (20000 points) and $t$, $\partial{M}/\partial\mu$ was computed numerically from equation \eqref{dmdmuex} with $n = 1.5$ ($\gamma = 5/3$), and the physical parameters were the same as those in Figure \ref{fig:rhoplots}. We see that the positions of the gas parcels disagree between the analytic solutions -- equations \eqref{rhosg} and \eqref{rhosh} -- and the numerical solutions that used the exact, directly computed forms for $\partial{r}/\partial\mu$ and $\partial{M}/\partial\mu$ (compare for example the locations at which the density of the bound material goes to zero between the left-hand panels of Figures \ref{fig:rhoplots} and \ref{fig:rhoplotsex}). In particular, the return time of the most bound material is vastly overestimated because of the incorrect scaling of the energies in the analytic case. However, the overall scaling of the density with $r$ is upheld. 

\begin{figure} 
   \centering
   \includegraphics[width=3.4in]{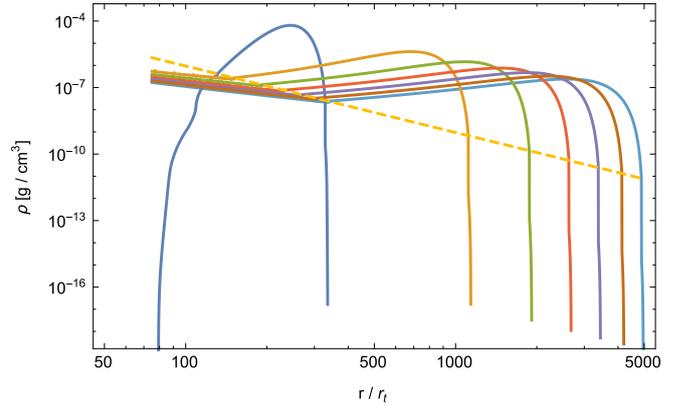} 
   \caption{The numerical solution for the density that adopts the piecewise behavior for $H$ (equation \ref{Hfull}) so that the density is shear dominated in regions where $\rho < M_h/(2\pi{r^3})$ and self-gravity dominated where $\rho > M_h/(2\pi{r^3})$, for the same parameters and times chosen in Figure \ref{fig:rhoplots}. The yellow, dashed line shows the curve $M_h/(2\pi{r^3})$.}
   \label{fig:rho_pw_plots}
\end{figure}

The solution that adopts a piecewise-continuous form for $H$ (equation \ref{Hfull}) is shown in Figure \ref{fig:rho_pw_plots}. The different curves correspond to the solution at the same times as Figures \ref{fig:rhoplots} and \ref{fig:rhoplotsex}, and the dashed line shows the curve $M_h/(2\pi{r^3})$. We see that these solutions possess properties of both the left and right-hand panels of Figure \ref{fig:rhoplotsex}: when the stream is self-gravity dominated, the density retains an inverted profile and increases as we move outward in specific energy. However, when the bound material nears the black hole and becomes shear-dominated, the tidal compression serves to decrease the stream width and produces an ``up-turn'' in the density profile as we approach the hole.

\begin{figure} 
   \centering
   \includegraphics[width=3.4in]{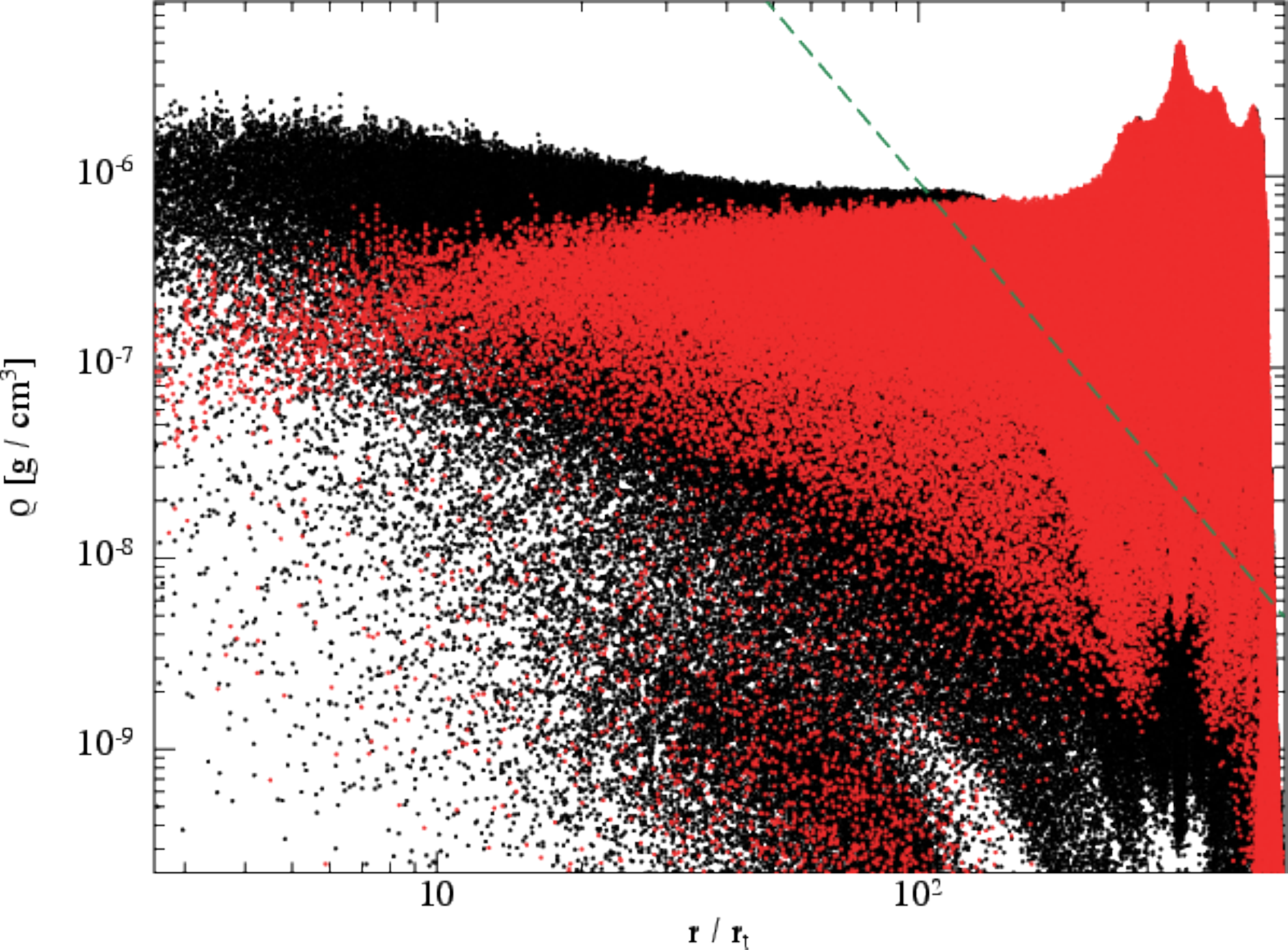} 
   \caption{A particle plot of the density in the stream $\rho$ in units of g / cm$^3$ as a function of $r$ in tidal radii at a time of 57 days post-disruption. The red points are from the simulation used in \citet{cou15a}, the black points are from an identical simulation but with $10^7$, as opposed to $10^6$, particles, and the green, dashed curve shows the function $M_h/(2\pi{r^3})$ -- the approximate dividing line between the shear and self-gravity dominated solutions. }
   \label{fig:rhocomps_particles}
\end{figure}

Interestingly, the density profile of the simulation in \citet{cou15a} does not exhibit this up-turn in the density as the material approaches the black hole. This anomalous behavior is demonstrated by the red points in Figure \ref{fig:rhocomps_particles}, which is a particle plot of density versus $r$ from the simulation in \citet{cou15a}. The black points in this Figure, on the other hand, are from a simulation that is identical to that from \citet{cou15a} except that, instead of $10^6$ particles, $10^7$ SPH particles were used. The green, dashed line shows the curve $M_h/(2\pi{r^3})$. The fact that the black points show the upturn in the density as we approach the black hole, consistent with what we expect from the analytic arguments presented here, shows that the disagreement between the simulation in \citet{cou15a} and our predictions arises from the resolution of the simulation. Since the increase in the density arises from the tidal compression exerted by the black hole, which results from the differential gravitational force acting across the diameter of the stream, we see that the width of the incoming debris stream is underresolved when only $10^6$ particles are used.

\section{Density scalings and fragmentation conditions}
The most accurate way of determining how the density of the stellar debris stream produced from a TDE varies with $r$ and $t$ is to solve equation \eqref{rho00} numerically. However, this direct way doesn't yield an obvious trend for how the density roughly scales with distance and time. Even from equations \eqref{rhosg} and \eqref{rhosh}, which give approximate expressions for $\rho$ in the self-gravity and shear-dominated limits, respectively, it is not overly apparent how the density qualitatively behaves as the stream recedes from the black hole. We can, however, discern how certain \emph{parts} of the stream behave by returning to equation \eqref{rho00} and focusing only on certain energy, or $\mu$, ranges. 

\subsection{Marginally bound material}
Recall that the marginally-bound segment of the stream corresponds to $\mu = 0$, follows $r \propto t^{2/3}$, and, from Section 2.2, satisfies $\partial{r}/\partial{\mu} \propto t^{4/3}$. Returning to equation \eqref{rho00}, we thus find

\begin{equation}
H^2\rho_m\propto t^{-4/3},
\end{equation}
where $\rho_m$ refers to the density at the marginally-bound part of the stream. If we now assume that the stream is self-gravity dominated, then using equation \eqref{Hsg} for $H$ shows

\begin{equation}
\rho_{m} \propto S^{-\frac{1}{\gamma-1}}t^{-\frac{4}{3(\gamma-1)}} \label{rhomsg}.
\end{equation}
If the gas is isentropic, then, for adiabatic indices of $\gamma = 1.5,$ 5/3, 1.8, and 2, we find that $\rho_{m} \propto t^{-8/3}$, $t^{-2}$, $t^{-5/3}$, and $t^{-4/3}$, respectively, which is in good agreement with the scalings found by \citet{cou16}. 

Equation \eqref{rhomsg} holds when the self-gravity of the debris dominates over the tidal field of the hole; however, if the density of the debris falls off at a rate steeper than $t^{-2}$, then even if it starts out as being self-gravitating, the tidal shear of the black hole will eventually dominate the self-gravity of the debris. As we saw above, the critical $\gamma$ at which self-gravity and the tidal field of the hole balance one another at the marginally-bound portion of the stream, which we will denote $\gamma_{c,m}$, is given by

\begin{equation}
\gamma_{c,m} = \frac{5}{3}.
\end{equation}
When the tidal shear dominates the self-gravity of the stream, the ability of the debris to collapse locally and fragment is seriously hindered. Therefore, we expect $\gamma_{c,m}$ to represent the critical adiabatic index at which fragmentation is marginally possible: for $\gamma > \gamma_{c,m}$, the stream can fragment, while for $\gamma < \gamma_{c,m}$, the stream is stable to fragmentation. This interpretation is supported by the results of \citet{cou16}, who found that the stream fragmented vigorously for $\gamma = 1.8$ and 2, fragmented at late times and due to numerical noise for $\gamma = 5/3$, and did not exhibit any fragmentation for $\gamma = 1.5$ (see also Section 6 for a further discussion of this point).

When the tidal shear of the black hole overwhelms the stream self-gravity, then using the fact that $H^2\propto{r^2}$ shows that the isentropic density scales as 

\begin{equation}
\rho_{m} \propto t^{-8/3}. \label{rhomsh}
\end{equation}
This demonstrates that the density of the stream will not necessarily fall off as a single power-law throughout its entire evolution. For example, if the stream starts out self-gravitating and has an adiabatic index of $\gamma = 1.5$, then it will initially follow $\rho_{m} \propto t^{-2.66}$ before eventually transitioning to the shallower power-law $\rho_{m} \propto t^{-2.22}$.

If we let the entropy scale as $S \propto t^{-p}$, which provides some insight into the ability of heating and cooling to affect the evolution of the stream density, then it is simple to show that the critical adiabatic index along the marginally-bound portion of the stream becomes

\begin{equation}
\gamma_{c,m} = \frac{5}{3}-\frac{p}{2}. \label{gammacm}
\end{equation}
This illustrates that cooling does not need to be very efficient to significantly alter the fragmentation properties of the stream. Alternatively, if shocks play a significant role in heating the gas, then the ability of the stream to fragment will be correspondingly diminished. On the other hand, once the gas temperature drops to about $10^4$K, the gas can start to recombine and heat the debris, stalling its temperature at $10^4$K and resulting in an adiabatic index of order unity. However, once the neutral fraction becomes significant, the optical depth decreases dramatically and recombinations will serve to cool the gas, corresponding to a larger effective $\gamma$ \citep{kas10}. 

\subsection{Unbound material}
For the material in the unbound portion of the stream, the gas parcels follow $r \propto t^{2/3}$ initially, but eventually transition to $r\propto t$. We therefore have $\partial{r}/\partial{\mu} \propto t$ when the material recedes to large distances from the hole, and we find that equation \eqref{rho00} becomes

\begin{equation}
H^2\rho_u \propto t^{-1} \label{rhou},
\end{equation}
where $\rho_u$ refers to the density in the unbound portion of the stream. If we now focus on the self-gravity dominated limit, using equation \eqref{Hsg} shows

\begin{equation}
\rho_u \propto t^{-1/(\gamma-1)}.
\end{equation}
From this expression we see that the density of the unbound material falls off at a \emph{shallower} rate than the marginally-bound debris. Specifically, for $\gamma = 5/3$ we find $\rho_u \propto t^{-3/2}$ as compared to $\rho_{mb} \propto t^{-2}$. Furthermore, since the density of the black hole falls off as $M_{h} \propto t^{-3}$ for the unbound debris as opposed to $M_h \propto t^{-2}$ for the marginally-bound debris, the critical adiabatic index at which the black hole density equals the stream density is correspondingly smaller. In particular, if we set $\rho_u \propto t^{-3}$, we find that the critical adiabatic index in the unbound portion of the stream is $\gamma_{c,u} = 4/3$. 

When the tidal shear of the hole dominates the self-gravity, using equation \eqref{Hex} in equation \eqref{rhou} gives

\begin{equation}
\rho_u \propto t^{-3}.
\end{equation}
As was true for the marginally-bound case, this shows that the density of the unbound material will not necessarily fall off as a single power-law throughout its evolution if it passes from being self-gravity to being shear-dominated. 

Furthermore, because the energies of the debris are greater than zero by only a small amount, the initial evolution of even the unbound debris will still follow approximately $r_{u} \propto t^{2/3}$, and only at some time $t_{tr}$ will the orbits transition to $r_u \propto t$. If we rewrite equation \eqref{rmuapp} as

\begin{equation}
r(\mu,t) \simeq \left(\frac{3}{2}\sqrt{2GM_h}t\right)^{2/3}+\frac{R_*\mu}{t_m^{4/3}}{t}^{4/3}, \label{rofmut}
\end{equation}
where the second term in this equation comes from equation \eqref{r1app} with the requirement that $r_i(t_m) = r_m(t_m)+\mu_iR_*$, then we see that the time at which this transition occurs is approximately

\begin{equation}
t_{tr} \simeq \frac{5\sqrt{5}}{3}\frac{q}{\mu^{3/2}}\frac{R_*^{3/2}}{\sqrt{2GM_h}}.
\end{equation}
Thus, if the unbound stream is self-gravitating, then at $t_{tr}$ we expect the density to transition from $\rho_u \propto t^{-2}$ to $\rho_u \propto t^{-1.5}$. Equivalently, if we are following an unbound fluid parcel with a narrow range of specific energies, then we expect the density to transition from $\rho_u \propto r^{-3}$ to $\rho_u \propto r^{-1.5}$ at this time. For the disruption of a solar-type star by a $10^6M_{\astrosun}$ SMBH, this time corresponds to $t_{tr} \simeq 48$ days post-disruption.

\begin{figure*} 
   \centering
   \includegraphics[width=0.49\textwidth]{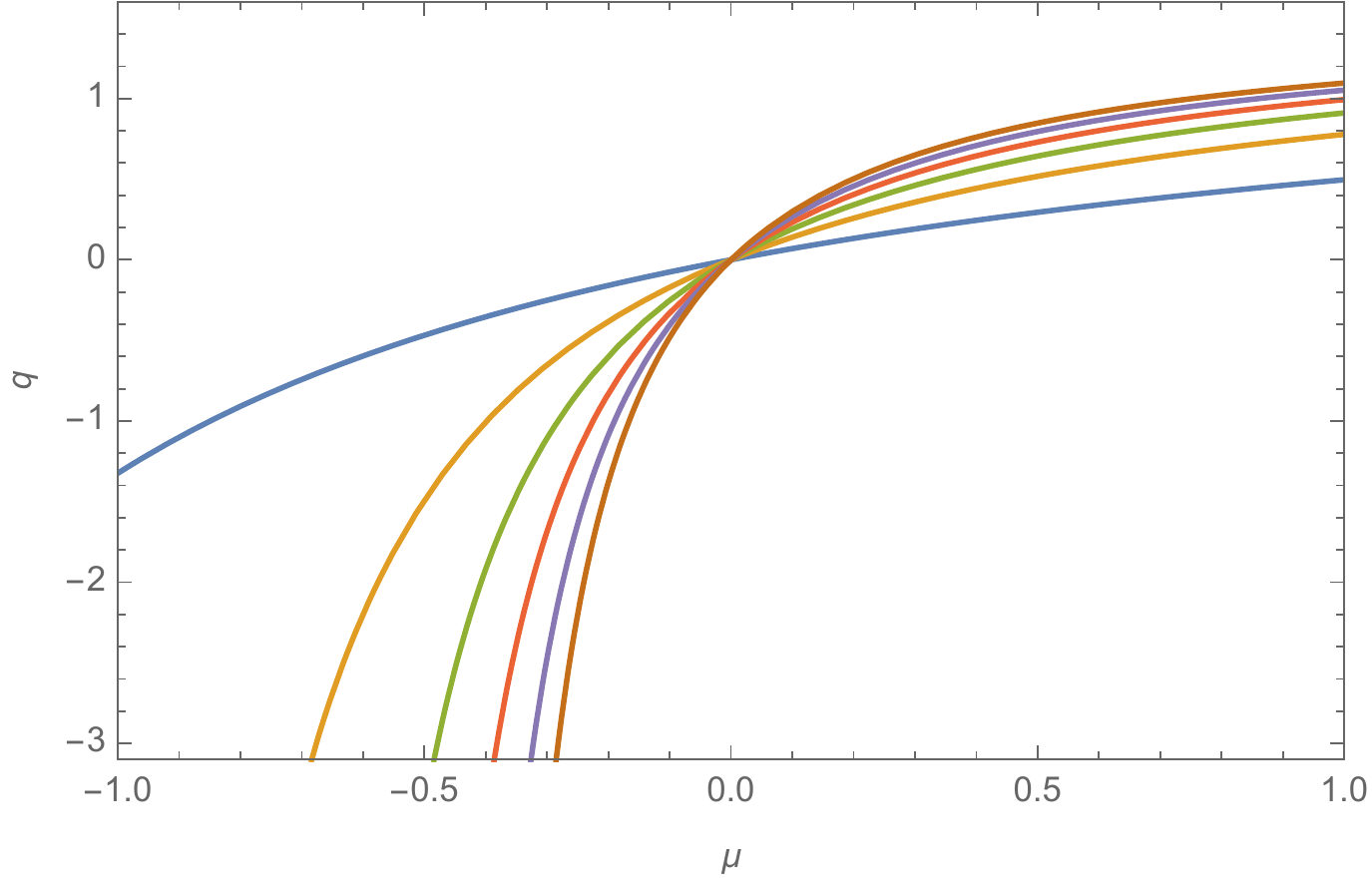}
   \includegraphics[width=0.49\textwidth]{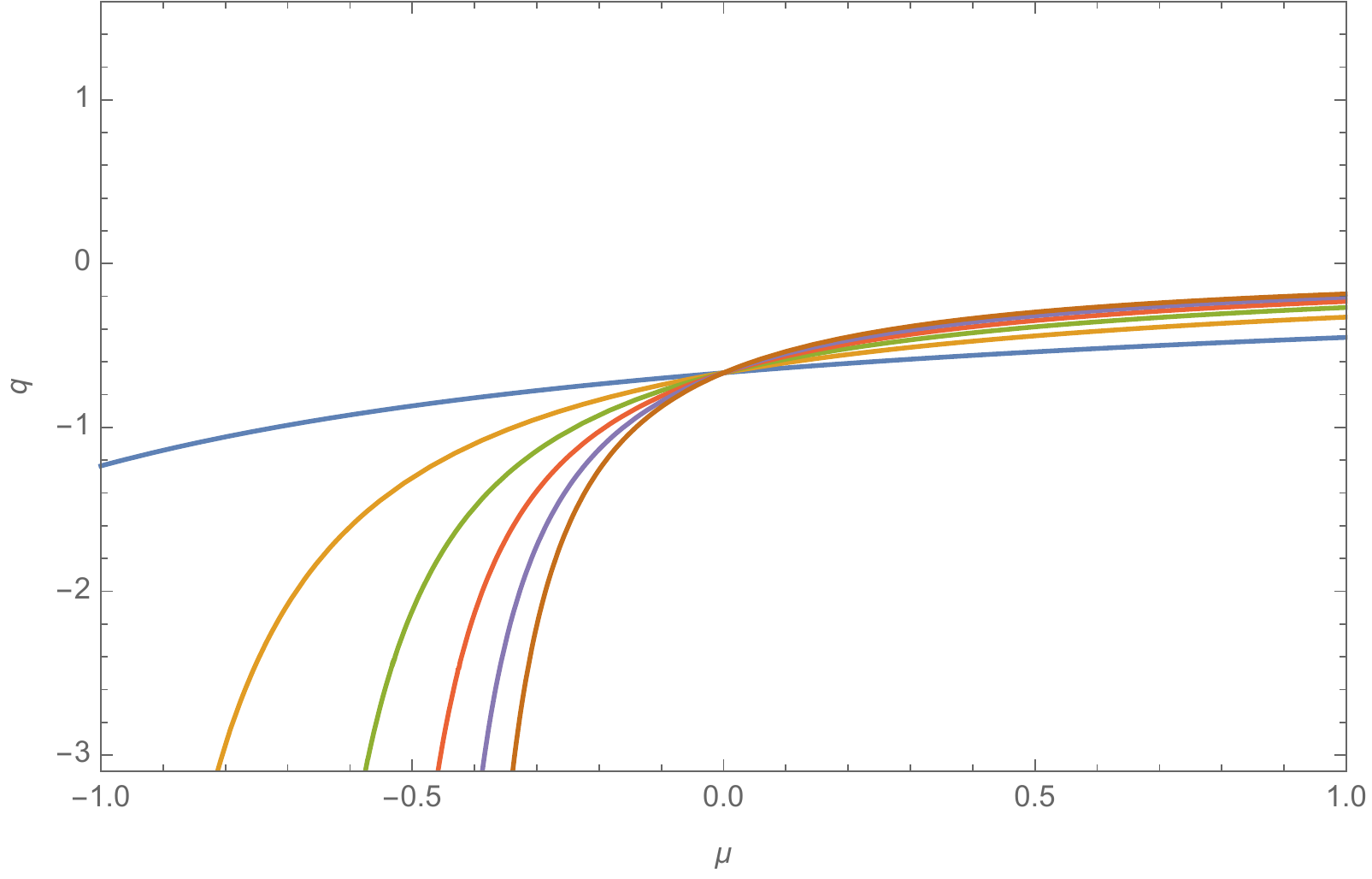}
   \caption{The quantity $q =d\ln(\rho{r^3})/d\ln{t}$ as a function of $\mu$ when $\gamma = 5/3$, $M_h = 10^6M_{\odot}$, the progenitor is solar, and the gas is isentropic; the flow is assumed to be self-gravity dominated in the left-hand panel and shear-dominated in the right-hand panel. The different curves show different times, with the blue curve at $1000\times{r_t^{3/2}}\sqrt{2GM_h} \simeq 13$ days after disruption, the brown curve at $11000\times{r_t^{3/2}}\sqrt{2GM_h} \simeq 143$ days after disruption, and each curve differs from the next closest curve by $2000\times{r_t^{3/2}}\sqrt{2GM_h} \simeq 26$ days (e.g., the yellow curve is 26 days after the blue curve, the green curve is 26 days after the yellow curve, etc.). The range of the $y$-axis was set to the same values for each plot for ease of comparison. Fluid elements in the stream follow vertical lines from one curve to the next.}
   \label{fig:qofmu}
\end{figure*}

\subsection{Bound material and overall scalings}
The above two subsections show that the marginally-bound and unbound material behave quite differently in terms of their asymptotic scalings. As was true for the unbound material, the initial evolution of the bound material will follow approximately $r_{b} \propto t^{2/3}$ because the specific energies, while being negative, are very close to zero. Since the bound material inevitably falls back to the origin no matter how close the energy is to zero, though, there is unfortunately no simple way of discerning the asymptotic behavior of this segment of the stream. 

One quantity that gives some insight into its behavior, however, is the number $q \equiv d\ln(\rho{r^3})/d\ln{t}$ as a function of $\mu$. When taking this logarithmic derivative, the quantities $\rho$ and $r$ are considered functions of $\mu$ and $t$ and $\mu$ is a constant. Therefore, $q$ gives the power-law index of the ratio of the density along the stream relative to the density associated with the black hole tidal field as a function of time and as we move along with a fluid element. 

Figure \ref{fig:qofmu} shows a plot of $q(\mu)$ for $\gamma = 5/3$ and the disruption of a solar-type star by a $10^6M_{\astrosun}$ SMBH ($S$ was set to a constant). The left-hand panel gives the solution when the flow is dominated by self-gravity, while the right-hand panel shows the case when the flow is shear dominated. The different curves give different times since disruption, with the blue curve at $ \simeq 13$ days after disruption, the brown curve at $ \simeq 143$ days after disruption, and each curve differing from the next closest curve by $ \simeq 26$ days. As we mentioned, the quantity $q$ characterizes the instantaneous power law of the product $\rho{r^3}$, i.e., the ratio of the density in the stream to that of the black hole scales instantaneously as $\rho{r^3} \propto t^{q}$. 

These plots show us what to expect when we consider lines of constant $\mu$ -- when we follow along a single fluid element. When $\mu > 0$, the specific energy of the gas parcel is greater than zero, and thus initially follow $r \propto t^{2/3}$ but eventually transition to $r \propto t$ for late times. We would thus expect that, at early times, the ratio of the density of the stream to that of the black hole should follow $\rho\,{r}^3 \propto 1$ ($\rho\,{r^3} \propto t^{-2/3}$) for self-gravity dominated (shear dominated) streams, while at late times it should follow $\rho\,{r}^3 \propto t^{1.5}$ ($\rho\,{r^3} \propto 1$) for self-gravity dominated (shear dominated) streams. This behavior is exactly shown in the left and right-hand panels of Figure \ref{fig:qofmu}, and specifically the rate at which this transition occurs. Likewise, if the material is \emph{exactly} at $\mu = 0$, it will always follow $\rho\,{r^3}\propto 1$ ($\rho\,{r^3}\propto t^{-2/3}$) for self-gravity (shear)-dominated streams, and this is seen in the Figure.

Figure \ref{fig:qofmu} also shows how the density of the bound material, fluid elements with $\mu < 0$, evolves with respect to the density of the SMBH. As we mentioned, there is no asymptotic behavior for the bound material because it inevitably falls back to the black hole. However, we see that the ratio of the stream density to the black hole density always scales as some power of $t$ that is less than 1, and in fact this power-law decreases as time increases. We thus expect that all material bound to the black hole, if it follows a $\gamma = 5/3$ equation of state, will eventually become shear-dominated. 

\section{Discussion}
In the preceding sections we developed pseudo-analytical expressions for the velocity profile of the debris, the positions of the gas parcels comprising the debris stream, and the density of the debris stream. Here we discuss how the angular momentum of the debris might alter these expressions, make comparisons to the previous work of \citet{koc94} on tidally-disrupted debris streams, consider the fate of the debris that satisfies the critical adiabatic index condition, and explore how the entropy of the gas should realistically behave.

\subsection{The neglect of angular momentum}
The solutions we have given here rely on the fact that the angular momentum of the material is small enough to be ignored when considering the evolution of the debris stream. Since the $\phi$-component of the velocity is approximately given by $r\dot\phi \simeq \ell/r$, where $\ell \simeq \sqrt{GM_h{r_p}}$ is the specific angular momentum of the material and $r_p \simeq r_t$ is the pericenter distance of the star, we find that $v_{\phi}/v_r \simeq (r/r_p)^{-1/2}$.  Thus, the neglect of angular momentum only breaks down for small radii, and should be increasingly accurate as the pericenter distance of the progenitor gets smaller. This argument also agrees with Figure \ref{fig:deltavplots}, which shows that the agreement between the self-similar solution and the numerical solution gets better as we proceed to larger distances from the hole (and, as a consequence, the ratio of the angular velocity to the radial velocity gets smaller). 

One consequence of accounting for the finite angular momentum of the material is that the position of the center of the stream (in terms of its width) is not purely radial, but instead sweeps out some angle (see also \citealt{gui15}). Thus, the extent of the stream is not exactly $(\partial{r}/\partial\mu)dr$, but is instead given by

\begin{equation}
\frac{\partial{r}}{\partial\mu}d\mu \rightarrow \sqrt{\left(\frac{\partial{r}}{\partial\mu}\right)^2+r^2\left(\frac{\partial\phi}{\partial\mu}\right)^2}d\mu,
\end{equation}
where $\phi$ is the in-plane angular position of the center of the stream; this is the equation employed by \citet{cou15a} and \citet{cou16}. Therefore, the overall length of the stream is slightly underestimated by ignoring the angular momentum of the gas.

A more important consequence of the angular momentum of the material is that the orientation of the tidal force of the SMBH will change as the bound material returns to the origin. Specifically, by assuming that the motion of the gas parcels was radial with a transverse extent $s$, the tidal field of the hole only served to compress material in the transverse direction (viz.~equation \ref{bhshear}) and increase the density of the stream. However, once the orientation of the cross-sectional width of the stream $H$ rotates to the point where it is aligned with the radial displacement of the center of the stream, the tidal compression will transition to a tidal shear that results in a decrease of the density.

\begin{figure} 
   \centering
   \includegraphics[width=2in]{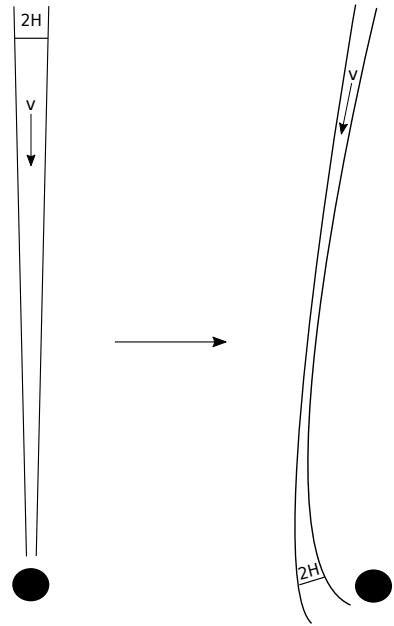} 
   \caption{A schematic of the debris stream, not drawn to scale, returning to the black hole under the approximations set out in our analysis (left drawing) and a more realistic distribution that accounts for the finite angular momentum of the debris (right drawing). This shows that, as the material gets very close to the hole, the tidal compression becomes a tidal shear that serves to decrease the density.}
   \label{fig:tidal_compression}
\end{figure}

Figure \ref{fig:tidal_compression} shows a rough drawing, not to scale, of the cross-section of the returning debris stream that illustrates this point: the left side shows the current assumption about the nature of $H$ versus the position of the center of the stream, while the right side gives a more accurate picture that accounts for the finite angular momentum of the material. From this schematic we see that, under the current, radial approximation for the position of the center of the stream (in terms of the transverse width), the tidal field of the black hole will only compress the stream  and consequently enhance the density near the black hole. In actuality, however, the small amount of angular momentum possessed by the debris causes it to miss the black hole, changing the tidal compression of the hole into a tidal shear that actually decreases the density.

\subsection{Comparison with Kochanek (1994)}
\citet{koc94} also performed an analysis of the debris streams produced from TDEs, in an attempt to discern the nature of the self-intersection that ensues when the bound material is swung through its general relativistic apsidal precession angle (roughly ten degrees for a solar-like star disrupted by a $10^6M_{\odot}$ SMBH; \citealt{ree88}). Here we make a brief comparison between his findings and ours. 

One important result of \citet{koc94} was that, if the impact parameter $\beta \equiv r_t/r_p$, where $r_p$ is the pericenter distance of the center of mass of the progenitor star, is large enough, then self-gravity will not be important for determining the evolution of the stream. Here we find a similar trend: from Figure \ref{fig:qofmu}, shear-dominated streams always follow $\rho\,{r^3} \propto t^{-q}$ with $q < 0$. Thus, if the impact parameter is high enough such that the density everywhere satisfies $2\pi\rho < M_h/r^3$, then the stream density will never be able to surmount the tidal shear of the SMBH (although the unbound material does asymptotically follow $\rho \propto r^{-3}$). 

\citet{koc94} also found that the width of the stream varies as $H \propto r^{1/4}$ for a $\gamma = 5/3$ equation of state. This result ultimately comes from the fact that the mass per unit length $\Lambda \equiv \partial{}M/\partial{}r$ is assumed to be $\Lambda \propto r^{-1}$ (see also \citealt{gui14} for the expression when general $\gamma$ are used alongside this prescription). However, we found from our analysis that the early evolution of the stream -- when all of the gas parcels are following approximately $r \propto t^{2/3}$ orbits -- is characterized by

\begin{equation}
\Lambda = \frac{\partial{}M}{\partial{}r} = \left(\frac{\partial{r}}{\partial\mu}\right)^{-1}\frac{\partial{M}}{\partial\mu} \propto t^{-4/3} \propto r^{-2},
\end{equation}
where the last relation follows from the fact that $r \propto t^{2/3}$. Since $H^2\rho \propto dM/dr$, and $\rho \propto t^{-2} \propto r^{-3}$ during this phase, we thus find

\begin{equation}
H \propto r^{1/2}, \label{Hofr}
\end{equation}
which disagrees with the scaling $H \propto r^{1/4}$ found by \citet{koc94}. For general $\gamma$, using equations \eqref{Hsg} and \eqref{rhomsg} gives

\begin{equation}
H \propto r^{\frac{2-\gamma}{\gamma-1}}.
\end{equation}

\begin{figure}
   \centering
   \includegraphics[width=3.4in]{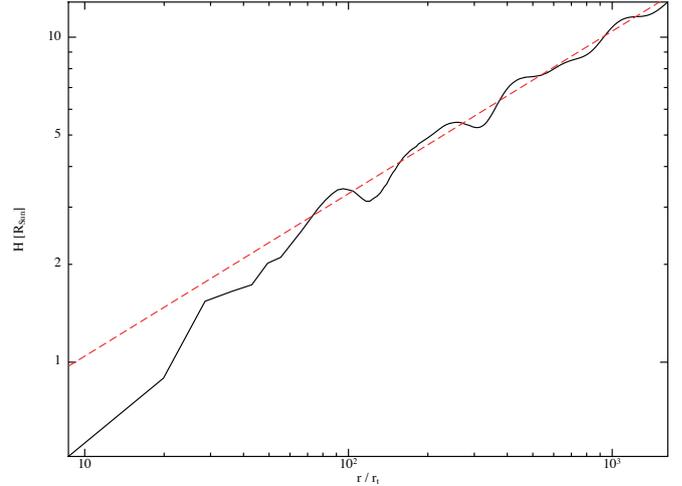} 
   \caption{{}{A log-log plot of the stream width, $H$, in units of solar radii at the marginally-bound location along the stream as a function of Lagrangian distance (i.e., moving with the marginally-bound portion of the stream) from the SMBH. The black, solid curve shows the solution from the same simulation performed in \citet{cou15a} (i.e., a solar-like star destroyed by a $10^6M_{\odot}$ SMBH) but with $10^7$ particles, while the red, dashed curve shows the analytic prediction $H/R_{\odot} \propto (r/r_t)^{1/2}$ (the constant of proportionality was set to 0.33). }}
   \label{fig:HvsR}
\end{figure}

{}{Figure \ref{fig:HvsR} shows $H/R_{\odot}$ at the marginally-bound portion of the stream as a function of $r/r_t$, with the solid, black curve from the simulation in \citet{cou15a} but using $10^7$ particles, and the red, dashed curve from the analytic prediction $H/R_{\odot} \propto (r/r_t)^{1/2}$ (the proportionality constant was set to 0.33, which provides a good by-eye fit). The width was calculated from the simulation by first determining the number of particles that had specific energies within a small range centered around zero; for this plot, the absolute value of the maximum energy was $10^{-6}c^2$ (as compared to the energy of the most energetic particles of $\sim2\times10^{-4}c^2$; see equation \ref{eps}), which amounted to roughly $10^5$ particles and a total mass of $\delta{}m \simeq .01M_{\odot}$ contained within the zero-energy bin. The maximum and minimum radii, $r_{max}$ and $r_{min}$, respectively, within that bin were computed, giving a radial extent $\delta{}r = r_{max}-r_{min}$ of the marginally-bound portion of the stream. We then calculated the quantity $dM/dr \simeq \delta{M}/\delta{r}$, determined the average density, and used equation \eqref{rho00} to calculate the width. This Figure shows that the analytic prediction of $H\propto r^{1/2}$ fits the data well. The disagreement between the prediction and the numerical results at early times is likely because self-gravity is still acting to modify the specific energies in the radial direction. This interpretation is supported by the fact that the number of particles in the marginally-bound energy bin goes from $\sim 8.2\times10^4$ to $\sim1.08\times 10^5$ during that initial transient phase, after which point it remains almost exactly constant. }

Once the orbits of the unbound material transition to $r \propto t$, which occurs around 50 days for the disruption of a solar-like star by a $10^6M_{\odot}$ SMBH, we recover the scaling $\partial{M}/\partial{r} \propto r^{-1}$. Using the fact that $\rho \propto t^{-3/2} \propto r^{-3/2}$ during this phase, the relation $H^2\rho \propto \partial{M}/\partial{r}$ yields $H \propto r^{1/4}$, which is in agreement with \citet{koc94}. 

\citet{koc94} also discussed how ionizations and recombinations, viscosity, and an ambient medium affect the evolution of the debris stream. We will return to these points in Section 6.4, where we reconsider the entropy of the gas.

\subsection{To fragment or not to fragment?}
When the density of the black hole $\rho_h \sim M_h/r^3$ falls off at a rate that is steeper than the density of the debris stream, we expect that the self-gravity of the stream in the radial direction will eventually overcome the shear of the black hole. Equivalently, if $\rho \gtrsim \rho_h$, the dynamical timescale $\tau_{d} \sim r^{3/2}/\sqrt{GM_h}$, which is the time over which material is sheared in the radial direction, will be longer than the free-fall timescale $\tau_{ff} \simeq 1/\sqrt{4\pi{G}\rho}$. Thus, in this case the material is able to aggregate faster than the rate at which it is being torn apart, and we expect the stream to be gravitationally unstable. Conversely, when $\rho_{}$ falls off more steeply than $r^{-3}$, the ordering of the timescales reverses, and overdensities are sheared apart faster than they can collapse. We thus expect the stream to be gravitationally stable when $\rho_{}$ falls off faster than $r^{-3}$.

These arguments are supported by \citet{cou16}, who investigated, from a numerical perspective, how the evolution of the debris stream from a TDE depends on the adiabatic index of the gas. They found that, when $\gamma = 1.8$ and 2, the stream fragmented early in its evolution; from our analysis $\rho_m \propto t^{-5/3}$ and $t^{-4/3}$ when $\gamma = 1.8$ and 2, respectively, and thus we expect the stream to fragment in these cases. On the other hand, \citet{cou16} found that the stream never fragmented -- even after ten years post-disruption -- when the adiabatic index of the gas was 1.5. From equations \eqref{rhomsg} and \eqref{rhomsh}, the density should fall off as $\rho \propto t^{-8/3}$ for $\gamma = 1.5$, which is significantly steeper than $t^{-2}$ and we therefore expect, consistent with the numerical findings, that the stream is stable to fragmentation in this case.

\citet{cou15a} and \citet{cou16} also ran simulations in which the adiabatic index of the gas was set to $\gamma = 5/3$. In this case, they found that the stream still fragmented, but the time at which fragmentation occurred depended on the resolution of the simulation. In particular, if the number of particles was increased, the stream fragmented later and vice versa. This finding then \emph{suggests} that the stream is gravitationally unstable, but that the noise inherent in the numerical method is seeding the instability. 

Is this finding and interpretation consistent with our analysis here? As we saw above, $\gamma = 5/3$ marks the critical adiabatic index (for an isentropic equation of state) where $\rho \propto t^{-2}$ during the early evolution of the debris. In this case, then, the freefall and collapse timescales are proportional to one another, and it is unclear whether or not the stream \emph{should} fragment. 

To answer this question, we note that a precisely analogous situation is encountered when one considers Jeans collapse in the early Universe: as primordial overdensities form and try to collapse, they are continuously stretched apart by the fact that the background density scales as $\rho \propto t^{-2}$ (see, e.g., \citealt{col95}, p.~202). In this case, then, the freefall and dynamical timescales are comparable, and the fate of the overdensities becomes unclear. However, one can show that there are still growing, unstable modes, but instead of evolving as exponentials they only scale as power-laws in time. To understand the origin of this result, we can argue somewhat heuristically that the relation

\begin{equation}
\dot{\delta\rho_{+}} \simeq \frac{\delta\rho_{+}}{\tau_{ff}},
\end{equation} 
where $\delta\rho_{+}$ is the growing mode of the instability and $\tau_{ff}$ is the free-fall timescale, holds more generally than just when $\tau_{ff}$ is time-independent (and, as a consequence, the instability grows exponentially). Thus, if we write $\tau_{ff} \simeq t/\sqrt{G\rho_0t_0^2}$, where $t_0$ is a characteristic timescale and $\rho_0$ is a characteristic density, then it follows immediately that $\delta\rho_{+} \propto t^{\sqrt{G\rho_0t_0^2}}$ -- the overdensities grow as power-laws.

For a star being disrupted by a SMBH, the characteristic timescale is the dynamical time at the tidal radius, so $t_0 \simeq r_t^{3/2}/\sqrt{GM_h}$, while the characteristic density $\rho_0$ is the original stellar density. We thus expect that when $\gamma = 5/3$, the growing modes should scale roughly as

\begin{equation}
\delta\rho_{+} \propto t^{\sqrt{\alpha}}, \label{deltarho+}
\end{equation}
where $\alpha = \rho_*r_t^3/M_h$ is the ratio of the stellar density, $\rho_*$, to the density of the SMBH at the time of disruption, $M_h/r_t^3$. 

These arguments illustrate that when the density of the stream falls off as $\sim1/r^3$ -- which occurs for the early evolution of the marginally-bound material when $\beta \sim 1$ and $\gamma = 5/3$, and also occurs for the shear-dominated portion of the unbound stream for any $\gamma$ -- there are growing unstable modes. A more rigorous approach would be to perform a perturbation analysis on the fluid equations, and when this is done for Jeans collapse in the early Universe, the results of the heuristic argument are upheld \citep{col95}. Since the problem is identical here in terms of the scalings of the variables, it follows that the heuristic argument is also valid, and for completeness we have checked that equation \eqref{deltarho+} is recovered if one performs a perturbation analysis. (Technically, the result is that $\rho \propto t^{N\sqrt{\alpha}}$, where $N$ is a number of order unity.) 

We thus find that the results of \citet{cou15a} and \citet{cou16} are consistent with our analytical arguments: when $\gamma = 5/3$, the stream is unstable to fragmentation. However, because the dynamical timescale and collapse timescale are proportional in this case, the overdensities grow as power-laws instead of exponentials. This considerably weaker growth rate means that resolving the instability is much more difficult, and the noise in the density that is induced by the numerical method becomes the dominant contributor to the fluctuations that seed the instability. However, in the presence of a sufficiently dense ambient medium, the shear between the stream and that medium would provide a physical origin for the perturbations \citep{bon15}.

Before moving on, we note that the simulations of \citet{cou15a} and \citet{cou16} found that bound portions of the stream, specifically those very close to the marginally-bound orbit, also exhibited fragmentation. Based on Figure \ref{fig:qofmu}, we would not expect this, as $\rho\,{r}^3 \propto t^{q}$ with $q$ less than zero for the bound material. However, our treatment here of the radial velocity ignored the effects of self-gravity in the radial direction; it can be seen from Figure \ref{fig:rhocomps_particles}, though, that the denser material towards the marginally bound segment of the stream generates a gravitational force on the bound material that serves to counteract the gravitational pull of the SMBH. This additional force will serve to reduce the amount of shear present within the flow, correspondingly lengthening the dynamical time and conceivably allowing some parts of the bound material to collapse.

\subsection{A more realistic entropy prescription}
Most of the examples we considered here assumed that the constant-entropy nature of the polytrope was preserved throughout the evolution of the debris. However, our equation \eqref{Hsg} permits a radially and temporally dependent entropy, which, through equation \eqref{rho00}, would serve to alter the nature of the stream. In particular, if the entropy decreases due to cooling, the density is correspondingly increased, making it easier for the stream to overcome the $\gamma = 5/3$ marginally-unstable condition (equation \ref{gammacm}); likewise, if the internal energy of the gas increases, the width of the stream grows and results in the gravitational stability of the stream.

When the stream initially starts to expand, any cooling is initially controlled by the fact that the surface radiates approximately as a blackbody:

\begin{equation}
dL = \sigma{}T_{eff}(r,t)^4dA,
\end{equation}
where $\sigma = 5.67\times10^{-5}$ [cgs], $dA$ is the area of the emitting surface, and $T_{eff}$ is the effective temperature that is, in general, dependent on time and position along the stream. The assumption of a blackbody is justified by the fact that the optical depth across the width of the stream is 

\begin{equation}
\tau \simeq \rho\kappa{H} \simeq \kappa\rho_*R_*\left(\frac{r}{r_t}\right)^{-5/2} \simeq \kappa\rho_*R_*\left(\frac{t}{t_0}\right)^{-5/3}, \label{tauThomson}
\end{equation}
where $\kappa$ is the Rosseland mean opacity and $\kappa\rho_*{R_*} \gtrsim 2\times10^{10}$ for solar-like progenitors ($R_* = R_{\odot}$ and $\rho_* \simeq 1$ g / cm$^3$). The stream is therefore very optically thick when the stream is still highly ionized, which justifies the assumption of local thermodynamic equilibrium (but see below for when recombinations start to change the opacity). The area of the stream is simply $dA \simeq 2\pi{H}dr$, $dr$ being the infinitesimal distance along the stream. Since the total amount of internal energy contained in the stream is comparable to the amount contained in the original star, $E_{int} \simeq GM_{*}^2/R_{*} \simeq 10^{48}$ ergs for a solar-like progenitor, any radiation escaping from the photosphere should decrease the internal energy on roughly the Kelvin-Helmholtz timescale -- thousands to millions of years if the disrupted star is roughly solar. Thus, the early evolution of the debris stream should be very nearly adiabatic.

However, recombinations will start to occur as the temperature of the debris reaches roughly $10^4$ K. As recognized by \citet{koc94} and \citet{kas10}, this could have important consequences for the evolution of the debris stream, heating the material when it is optically thick and cooling it as it becomes optically thin. If the stream follows a $\gamma = 5/3$ adiabatic equation of state and is self-gravity dominated, then we can show, by returning to equation \eqref{rho00}, using equations \eqref{Hsg}, \eqref{rofmut} and \eqref{dmdmuex}, and the ideal gas law, that the temperature evolves as

\begin{equation}
T \simeq 2.35\frac{GM_*m}{k_BR_*}\left(\frac{t}{t_m}\right)^{-4/3},
\end{equation}
where the factor of $2.35$ resulted from evaluating equation \eqref{dmdmuex} numerically for n=1.5 ($\gamma = 5/3$) at $\mu = 0$, $m \simeq 1.67\times10^{-24}$ g is the mean molecular mass in the stream, $k_B = 1.38\times10^{-16}$ [cgs] is Boltzmann's constant, and $t_m = (\sqrt{5}r_t)^{3/2}/\sqrt{2GM_h} \simeq 1.05$ h for a solar-like progenitor and a $10^6 M_{\odot} $ SMBH. Solving this equation for $t_{rec}$, the time at which $T = 10^4$, gives

\begin{equation}
t_{rec} \simeq 27\text{ days} \label{trec}
\end{equation}
after disruption. This number could be slightly sooner or later depending on the mean molecular weight (we adopted $m = m_h = 1.67\times10^{-24}$ g) and the initial temperature of the star. 

Equation \eqref{trec} should really be interpreted as an average value over the entire extent of the stream. In particular, since the density, and hence the temperature, is maximized near the marginally-bound portion of the stream (see Figure \ref{fig:rhoplots}), recombinations will start to occur at a later time than they will toward the radial extremities of the stream where the temperature is correspondingly lower. Likewise, the stream is not an infinitely thin line, but has a finite width $H$; near $H$ the gas temperature is lower, generating an earlier recombination time.

Initially -- when the gas is still largely ionized -- the optical depth is quite high, as evidenced from equation \eqref{tauThomson}, meaning that recombinations will heat the gas and keep it at a roughly constant temperature of $10^4$ K. Once recombinations start to occur, however, the decrease in the ionized fraction will correspondingly lower the opacity of the gas and allow more of the stream to cool, generating a ``cooling front'' that moves inward from $H$ \citep{kas10}.  Thus, while recombinations will initially heat the gas and serve to increase the specific entropy $S$, at later times the optical thinness of the stream means that continued recombinations will actually serve to decrease the specific entropy.

The local shear within the stream could also serve to heat the gas. If we prescribe the coefficient of dynamic viscosity by $\eta$, then viscous heating will modify the gas energy equation via \citep{cou14b}

\begin{equation}
\frac{p}{\gamma-1}\left(\frac{\partial}{\partial{t}}\ln{S}+v_r\frac{\partial}{\partial{r}}\ln{S}\right) \simeq \eta\left(\frac{\partial{v_r}}{\partial{r}}\right)^{2},
\end{equation}
where $p$ is the gas pressure and $S = p/\rho^{\gamma}$ is the entropy as we defined it for a polytrope. We see that the right-hand side of this equation falls off approximately as $\propto 1/r^3$ (assuming that the temporal and spatial dependence of $\eta$ is not too large), and consequently the viscous heating of the gas should not play a large role in the overall evolution of the debris stream.

Finally, the background radiation field present in the circumnuclear medium will also alter the thermodynamical evolution of the stream. If the stream is optically thin, then the entire stream would be maintained at the temperature of the radiation field $T_b$ (in the most extreme case, $T_b$ would be the temperature of the cosmic microwave background). When the stream is still optically thick, on the other hand, the radiation will be reprocessed by a thin, outer sheath \citep{koc94}.

When the recombinations or radiation from background sources start to heat the gas, it will become over-pressured with respect to self-gravity. In this case, if we return to equation \eqref{smom2}, we see that the cross-sectional width of the stream varies as

\begin{equation}
H \simeq c_st,
\end{equation}
where

\begin{equation}
c_s \simeq \sqrt{\frac{p}{\rho}}
\end{equation}
is the adiabatic sound speed. If we now use this expression for $H$ in equation \eqref{rho00} and again assume that the evolution proceeds adiabatically ($p = S\rho^{\gamma}$, where this $S$ is a constant that is larger than the value before heating started), then we find

\begin{equation}
\rho^{\gamma}t^2 \propto \left(\frac{\partial{r}}{\partial\mu}\right)^{-1}\frac{\partial{M}}{\partial\mu}.
\end{equation}
While the stream is still in the marginally-bound phase, $\partial{r}/\partial{\mu} \propto t^{-4/3}$, and we thus find

\begin{equation}
\rho \propto t^{-\frac{10}{3\gamma}}.
\end{equation}
In this case, we see that $\gamma = 5/3$ still marks the critical $\gamma$ at which $\rho \propto t^{-2} \propto r^{-3}$. 

\section{Summary and conclusions}
In this paper we have presented a semi-analytic analysis of the debris streams produced from TDEs, showing that there is a simple, self-similar prescription for the radial velocity of the debris. Specifically, we found that if one assumes that the radial velocity varies as $v_r = \sqrt{2GM_h/r}f(\xi)$, where $f$ is a function of the self-similar variable $\xi = \sqrt{2GM_h}t/r^{3/2}$, then the function $f$ is given by the solution to the equation

\begin{equation}
f' = \frac{f^2-1}{2-3f\xi},
\end{equation}
with the boundary condition $f(2/3) = 1$. Figure \ref{fig:vcomps} compares this velocity profile to the numerical solution of \citet{cou15a}, and shows that the two agree very well (see also Figure \ref{fig:deltavplots}). 

We also computed the cross-sectional radius of the stream, $H$, as a function of time and density. We found that, when the density and pressure are high enough, the stream width varies in a quasi-hydrostatic manner as $H^2 \propto \rho^{\gamma-2}$. However, if the shear of the black hole dominates self-gravity and pressure, then the entire stream evolves self-similarly with $H \propto r$.

The self-similar velocity profile and the solution for the cross-sectional radius of the stream allowed us, in conjunction with equation \eqref{rho00}, to solve for the density of the debris stream, $\rho$, as a function of space and time. Our approximate, analytical expressions (equations \ref{rhosg} and \ref{rhosh}) were shown to agree well with the more exact, numerical solutions that used the full forms for the quantities $\partial{r}/\partial\mu$ and $\partial{M}/\partial\mu$ (see Figures \ref{fig:rhoplots} and \ref{fig:rhoplotsex}). An inverted density profile -- where the density increases as a function of $r$ -- is actualized when the material is self-gravitating, while the density decreases monotonically when the shear of the black hole dominates the stream self-gravity.

We analyzed the general scaling of $\rho$ with time and distance from the black hole, and showed that, during the initial stage of evolution when the orbits of the gas parcels comprising the debris stream follow $r \propto t^{2/3}$, the portions of the stream that are self-gravitating satisfy $\rho \propto t^{-4/(3(\gamma-1))}$. This scaling demonstrates that $\gamma = 5/3$ marks the critical adiabatic index above which self-gravity always dominates the tidal shear of the black hole, below which the self-gravity of the stream becomes negligible. On the other hand, the unbound portion of the stream eventually follows $r \propto t$, which shows $\rho \propto t^{-1/(\gamma-1)}$ and the critical adiabatic index is $\gamma = 4/3$. Finally, when the stream falls below the self-gravitating limit (when $2\pi\rho < M_h/r^3$), the entire stream behaves self-similarly and $\rho \propto r^{-2}\left(\partial{r}/\partial\mu\right)^{-1}$. During this self-similar regime, the density falls off as $r^{-4}$ when $r \propto t^{2/3}$ (marginally-bound material, early evolution of the entire stream) and $r^{-3}$ when $r \propto t$ (unbound material, late evolution).

Finally, we showed that the finite angular momentum of the debris, which was ignored in our self-similar models, should have a small but noticeable effect as material falls back to the black hole. Comparisons between our results and those of  \citet{koc94} were made, demonstrating that we agreed on many aspects of the evolution of the debris stream but disagreed on some; in particular, our finding that the width of the stream scales as $H \propto r^{1/2}$ for a $\gamma = 5/3$, self-gravitating stream contrasts the scaling $H \propto r^{1/4}$ found by \citet{koc94}. We showed that when $\gamma = 5/3$, the stream of debris produced from a TDE should be gravitationally unstable; however, because the density of the stream scales as $\rho \propto r^{-3}$ for early times, which is identical to that of the black hole, the perturbations grow only as power-laws in time (specifically, $\delta\rho_{+} \propto t^{\sqrt{\alpha}}$, where $\alpha \simeq \rho_*r_t^3/M_h$ is the ratio of the density of the stellar progenitor to the density of the black hole at the time of disruption). We also considered how the entropy of the debris stream might evolve more realistically, given that the gas will start to recombine after a certain time and that the ambient medium may serve to heat the gas.

{}{One of the important conclusions of this analysis is that larger adiabatic indices provide enhanced gravitational fragmentation of the debris streams from TDEs, which is in agreement with numerical findings \citep{cou16}. Interestingly, this trend is in disagreement with those encountered in star formation, where larger adiabatic indices \emph{prevent} fragmentation \citep{lar03}. The reason for this discrepancy is that, as a gas cloud collapses under its own self-gravity, the density at any radius within the cloud is an increasing function of time. For an adiabatic equation of state, this increase in the density then corresponds to an increase in pressure, with larger adiabatic indices generating correspondingly larger increases in pressure. This increase in pressure then resists collapse to smaller scales. On the other hand, the density is a \emph{decreasing} function of time as the debris stream generated from a TDE recedes from the black hole, meaning that larger adiabatic indices generate more drastic decreases in pressure for the same decrease in density. It is due to this loss of pressure support that the trend of decreased stability for larger adiabatic indices is realized in tidally-disrupted debris streams.}

{}{The adiabatic indices of most main-sequence stars lie between 4/3 and 5/3 \citep{han04}. Based on the fact that our findings here suggest $\gamma_c = 5/3$ is the critical adiabatic index above which fragmentation is possible, one is led to the conclusion that most stars will not be susceptible to gravitational instability. However, it should be noted that the cooling front initiated by radiative recombination (see Section 6.4) could significantly alter the equation of state of the gas, causing the stream to transition from marginally-stable to unstable. Furthermore, the critical adiabatic index that demarcates the stability threshold is itself a function of both position along the stream and time, and falls to as low as 4/3 for the unbound portion of the stream at late times (see Section 5 and Figure \ref{fig:qofmu}). Thus, as long as the stream remains above the self-gravitating limit, the unbound portions of many tidally-disrupted debris streams should be gravitationally unstable. Finally, even though adiabatic indices greater than 5/3 are difficult to realize for most stars, values of 2 and greater are thought to be obtained for planets \citep{fab05} and compact objects \citep{lee07}. Planets tidally disrupted by their host stars and the remnants of compact object mergers should therefore be unstable to self-gravitational fragmentation.}

Our investigations assumed that the density of the circumnuclear medium does not affect the propagation of the debris stream produced from a TDE. Because the original stellar density is so much higher than the densities near the centers of most galaxies, this assumption is justified, with the bulk dynamics of the material largely unaffected \citep{koc94}. However, the Kelvin-Helmholtz instability could be important for modifying the density profile of some portions of the returning debris stream \citep{bon15}, and could also provide a physical length scale over which density perturbations occur that cause the stream to fragment. Also, once the stream propagates out to very large distances from the SMBH, the stream density will become comparable to that of the ambient medium, and drag effects will start to alter its motion as the stream enters a Sedov-Taylor phase \citep{gui15}.

In order to make comparisons with past simulations, we only computed the explicit velocity profile (the one that depends on both $r$ and $t$) and compared it to the numerical solution for the case when a solar-like star is destroyed by a $10^6M_{\odot}$ SMBH and the pericenter distance of the progenitor star was equal to the tidal radius. However, we expect that this self-similar solution is accurate over a very wide range of parameters (SMBH mass, stellar mass, impact parameter $\beta$, etc.) because it depends only on the local velocity and the local dynamical time, and this concern only for local quantities should be true for most TDEs (i.e., we expect that the specifics of the initial conditions of the encounter are rapidly forgotten as the stream recedes from the black hole). Even for encounters in which the star is only partially destroyed or recollapses to form a bound core \citep{gui13}, this self-similar velocity profile should hold for the disrupted debris and only break down once we get very near to the surviving core.

The analysis performed here was done with the specific application of stars disrupted by supermassive black holes in mind. However, we note that many of the results we obtained can be applied to other tidally-interacting systems. In particular, the self-similar velocity profile and the physical properties of the debris stream (the width and density distributions) described here are also applicable to compact object mergers (e.g., \citealt{lee07}), the disruptions of planets by stars (e.g., \citealt{fab05}), and conceivably the tidal tails generated from galactic interactions (e.g., \citealt{kni12}).

\section*{Acknowledgments}

This work was supported in part by NASA Astrophysics Theory Program grants NNX14AB37G and NNX14AB42G, NSF grant AST-1411879, and NASA's Fermi Guest Investigator Program. CN was supported by the Science and Technology Facilities Council (grant number ST/M005917/1). We used {\sc{splash}} \citep{pri07} for the visualization. Research in theoretical astrophysics at Leicester is supported by an STFC Consolidated Grant. This work utilized the Complexity HPC cluster at the University of Leicester which is part of the DiRAC2 national facility, jointly funded by STFC and the Large Facilities Capital Fund of BIS.

\bibliographystyle{mnras}

\begin{thebibliography}{}
\bibitem[Bogdanovic et al.(2014)]{bog14} Bogdanovic T., Cheng R.M., \& Amarao-Seoane P. 2014, ApJ, 788, 99
\bibitem[Bonnerot et al.(2015)]{bon15} Bonnerot, C., Rossi, E.M., \& Lodato, G. 2015, MNRAS, submitted (arXiv:1511.00400)
\bibitem[Burrows et al.(2011)]{bur11} Burrows D.N., Kennea J.A., Ghisellini G., et al. 2011, Nature, 476, 421
\bibitem[Cenko et al.(2012)]{cen12} Cenko S.B., Krimm H.A., Horesh A., et al. 2012, ApJ, 753, 77
\bibitem[Coles \& Lucchin(1995)]{col95} Coles, P. \& Lucchin, F. 1995, Cosmology. The Origin and Evolution of Cosmic Structure (England: John Wiley \& Sons)
\bibitem[Coughlin \& Begelman(2014a)]{cou14a} Coughlin E.R., \& Begelman, M.C. 2014a, ApJ, 781, 82
\bibitem[Coughlin \& Begelman(2014b)]{cou14b} Coughlin, E.R., \& Begelman, M.C. 2014b, ApJ, 797, 103
\bibitem[Coughlin \& Nixon(2015)]{cou15a} Coughlin E.R., \& Nixon C.J. 2015 ApJ, 808, L11
\bibitem[Coughlin et al.(2016)]{cou16} Coughlin, E. R., Nixon, C. J., Begelman, M. C., \& Armitage, P.J. 2016, MNRAS, 455, 3612
\bibitem[Evans \& Kochanek(1989)]{eva89} Evans C.R., \& Kochanek, C.S. 1989, ApJ, 346, L13
\bibitem[Faber et al.(2005)]{fab05} Faber, J.A., Rasio, F.A., \& Willems, B. 2005, Icar, 175, 248
\bibitem[Frank \& Rees(1976)]{fra76} Frank J., \& Rees, M.J. 1976, MNRAS, 176, 633
\bibitem[Gezari et al.(2008)]{gez08} Gezari, S., Basa S., Martin D.C., et al. 2008, ApJ, 676, 944
\bibitem[Hansen et al.(2004)]{han04} Hansen C.J., Kawaler S.D., Trimble V., 2004, Stellar Interiors (New York: Springer)
\bibitem[Guillochon \& Ramirez-Ruiz(2013)]{gui13} Guillochon, J. \& Ramirez-Ruiz, E. 2013, ApJ, 767, 25
\bibitem[Guillochon et al(2014)]{gui14} Guillochon, J., Manukian, H., \& Ramirez-Ruiz, E. 2014, 783, 23
\bibitem[Guillochon et al.(2015)]{gui15} Guillochon, J., McCourt, M., Chen, X., et al. 2015, ApJ, submitted (arXiv:1509.08916)
\bibitem[Kasen \& Ramirez-Ruiz(2010)]{kas10} Kasen, D., \& Ramirez-Ruiz, E. 2010, ApJ, 714, 155
\bibitem[Knierman et al.(2012)]{kni12} Knierman, K., Knezek, P.M., Scown, P., et al. 2012, ApJ, 749, L1
\bibitem[Kochanek(1994)]{koc94} Kochanek, C. S. 1994, ApJ, 422, 508
\bibitem[Komossa \& Bade(1999)]{kom99} Komossa S., \& Bade N. 1999, A\&A, 343, 775
\bibitem[Komossa(2015)]{kom15} Komossa, S. 2015, JHEAp, 7, 148
\bibitem[Lacy et al.(1982)]{lac82} Lacy J.H., Townes C.H., \& Hollenbach D.J. 1982, ApJ, 262, 120
\bibitem[Larson(2003)]{lar03} Larson, R.B. 2003, RpPh, 66, 1651
\bibitem[Lee \& Ramirez-Ruiz(2007)]{lee07} Lee, W.H., \& Ramirez-Ruiz, E. 2007, NJPh, 9, 17
\bibitem[Lodato et al.(2009)]{lod09} Lodato, G., King, A. R., \& Pringle, J. 2009, MNRAS, 392, 332
\bibitem[Miller et al.(2015)]{mil15} Miller, J.M., Kaastra, J.S., Miller, M.C., et al. 2015, Nature, 526, 542
\bibitem[Phinney(1989)]{phi89} Phinney, E. S. 1989, in IAU Symp. 136, The Center of the Galaxy, ed. M. Morris
(Cambridge: Cambridge Univ. Press), 543
\bibitem[Price(2007)]{pri07} Price, D.J. 2007, PASA, 24, 159
\bibitem[Price \& Federrath(2010)]{pri10} Price, D.J., \& Federrath, C. 2010, MNRAS, 406, 1659
\bibitem[Rees(1988)]{ree88} Rees M.J., 1988, Nature, 333, 523
\bibitem[Stone \& Metzger(2014)]{sto14} Stone, N.C., \& Metzger, B.D. 2016, MNRAS, 455, 859
\end{thebibliography}

\bsp	
\label{lastpage}
\end{document}